\documentclass[
superscriptaddress,
aps,prxquantum,
reprint,
twocolumn,
amsmath,amssymb,showpacs
]{revtex4-2}

\usepackage{amsmath,mathtools}
\usepackage{mathrsfs}
\usepackage{ulem}
\usepackage{changes}
\usepackage{amsfonts}
\usepackage{physics}
\usepackage{txfonts}
\usepackage{dsfont}
\usepackage{amssymb}
\usepackage{graphicx}
\usepackage{float}
\usepackage{xcolor} 
\usepackage{changes}
\usepackage{url}
\usepackage[colorlinks=true, citecolor=blue, urlcolor=blue]{hyperref}

\begin{document}

\title{Robust and digital hyper-polarization protocol of nuclear spins via
magic sequential sequence}

\author{Haiyang Li}
\affiliation{Faculty of Arts and Sciences, Beijing Normal University, Zhuhai 519087, China}

\author{Hao Liao}
\affiliation{National Engineering Laboratory on Big Data System Computing Technology, Guangdong Province Engineering Center of
China-made High Performance Data Computing System, College of Computer Science and Software Engineering, Shenzhen
University, Shenzhen 518060, China}

\author{Ping Wang}
\email{wpking@bnu.edu.cn}
\affiliation{Faculty of Arts and Sciences, Beijing Normal University, Zhuhai 519087, China}

\date{\today}

\begin{abstract}
Hyper-polarization of nuclear spins is crucial for advancing nuclear
magnetic resonance (NMR) and quantum information technologies, as
nuclear spins typically exhibit extremely low polarization at room
temperature due to their small gyro-magnetic ratios. A promising approach
to achieving high nuclear spin polarization is transferring the polarization
of electron to nuclear spin. The nitrogen-vacancy (NV) center in diamond
has emerged as a highly effective medium for this purpose, and various
hyper-polarization protocols have been developed. Among these, the
pulsed polarization (PulsePol) method has been extensively studied
due to its robustness against static energy shifts of the electron
spin. In this study, we introduce a sequential polarization protocol
and identify a series of magic and digital sequences for hyper-polarizing
nuclear spins. Notably, we demonstrate that some of these magic sequences
exhibit significantly greater robustness compared to the PulsePol
protocol in the presence of finite half $\pi$ pulse duration of the
protocol. This enhanced robustness positions our protocol as a more
suitable candidate for hyper-polarizing nuclear spins species with
large gyromagnetic ratios and also ensures better compatibility with
high-efficiency readout techniques at high magnetic fields. Additionally,
the generality of our protocol allows for its direct application to
other solid-state quantum systems beyond the NV center. 
\end{abstract}

\maketitle


\section{Introduction}

Enhancing nuclear magnetic resonance (NMR) signals is crucial for
a variety of applications, including biological research\cite{DumezAnalyst2015,Aslam2023},
drug discovery\cite{SternJPCL2015}, and nuclear spin-based gyroscope\cite{SoshenkoPRL2021}.
However, the utility of NMR is often constrained by the inherently
low Boltzmann polarization of nuclear spins, which, for protons at
room temperature, is typically on the order of $10^{-5}$. To address
this limitation, dynamic nuclear polarization (DNP) has been developed
as an effective strategy\cite{AbragamROPP1978,EillsChemRev2023}.
This technique leverages the high polarization of electron spins and
transfers it to nuclear spins to significantly enhance NMR sensitivity.

In recent years, the electron spins associated with nitrogen-vacancy
(NV) centers in diamond have gained attention as a promising medium
for DNP\cite{JacquesPRL2009,FischerPRB2013,GreenPRB2017,BroadwayNC2018,HenshawPNAS2019,DuartePRB2021,ChenPRB2024}.
NV centers are particularly advantageous due to their long coherence
times (on the order of milliseconds), high degree of optical polarization,
rapid optical polarization rates (in the microsecond range), and excellent
controllability at room temperature\cite{DohertyPR2013}. These properties
make NV centers an ideal candidate for advancing the field of nuclear
spin polarization and its applications.

A series of polarization techniques have been proposed and demonstrated
to polarize the surrounding nuclear spins in NV centers. These techniques
can be categorized into two groups: all-optical polarization methods
\cite{JacquesPRL2009,FischerPRB2013,GreenPRB2017,DuartePRB2021} and
microwave-assisted optical polarization methods \cite{HenstraJOMR1988,LondonPRL2013,ChenPRB2015,AlvarezNC2015,ScheuerNJP2016,SchwartzSA2018,AjoyPNAS2018,AjoySA2018,ZangaraPNAS2019,AjoyJOMR2021,HealeyPRAppl2021}.
The key challenge in polarization techniques lies in how to match
the large energy gap between the electron spin and nuclear spins.
In all-optical methods, this energy matching is typically achieved
by tuning the electron energy splitting to resonate with the nuclear
spin near the level anti-crossing (LAC) point \cite{JacquesPRL2009,FischerPRB2013,FischerPRL2013,WangNC2013,WangNJP2015,WangEPJD2016,BroadwayNC2018}.
However, the polarization behavior near the LAC point is often sensitive
to inhomogeneous broadening of the energy split, making it highly
dependent on magnetic field stability and energy disorder in ensemble
system. To address this limitation, microwave-assisted methods have
been developed in recent years. These methods achieve energy matching
through microwave driving or timing \cite{HenstraJOMR1988,LondonPRL2013,ChenPRB2015,AlvarezNC2015,ScheuerNJP2016,SchwartzSA2018,AjoySA2018,LangPRL2019,TanSA2019,ZangaraPNAS2019,HealeyPRAppl2021,AjoyJOMR2021,WiliSA2022,RedrouthuJACS2022}.
A significant advancement in this area is the development of pulse
polarization methods (PulsePol) based on Hamiltonian engineering \cite{SchwartzSA2018}.
This approach has been proven to be robust against inhomogeneous broadening
in ensemble NV centers \cite{SchwartzSA2018,HealeyPRAppl2021} and
has been widely investigated\cite{SasakiPRB2018,SasakiAPL2020,RandallScience2021,ShenPRL2023,SasakiPRL2024}.

Despite its advantages, the performance of the PulsePol method would
deteriorate under high magnetic fields due to the finite duration
of the $\pi/2$ pulses, which is typically constrained by the maximum
of available microwave driving power. Under this case, it is unclear
whether the PulsePol method still performs the best. Nevertheless,
high magnetic fields have obvious advantage that make them desirable
for polarization protocols, as demonstrated in numerous studies~\cite{KingPRB2010,TanSA2019,SahinNC2022}.
First, high magnetic fields provide greater chemical shift dispersion~\cite{SahinNC2022}.
Second, the robustness of polarization protocol to high magnetic field
can extend the applicability of polarization protocols to nuclear
species with large gyromagnetic ratios and enhances compatibility
with other high-field techniques in NV center, such as high-efficiency
readout methods~\cite{NeumannScience2010}, which are crucial for
NMR and quantum sensing applications. Finally, the prolonged lifetimes
of nuclear spin targets in high magnetic fields can significantly
improve polarization transfer efficiency to external nuclear spins
through spin diffusion processes~\cite{CheungPRB1981,FernandezNanoLett2018,ShagievaNanoLett2018,TetiennePRB2021,HealeyPRAppl2021}.

In this work, we propose novel hyper-polarization protocols that incorporate
three time delays to address the robustness limitations imposed by
finite $\pi/2$ pulse durations. We derive a family of magic sequences
that simultaneously maximize both the polarization degree and polarization
rate, with the PulsePol method emerging as a special case of our more
general protocol. We find that some magic sequences exhibit significantly
enhanced robustness to finite $\pi/2$ pulse durations compared to
the conventional PulsePol method, for both steady-state polarization
and polarization rate. These digital polarization sequences feature
well-defined, stable timing recipes, making them particularly convenient
for applications involving polarization transfer to both internal
and external nuclear spins in NV ensemble systems. Furthermore, we
obtain an analytical expression for the polarization rate which can
be used to measure precisely the transverse hyperfine coupling of
target nuclear spins.

\section{Protocol and theoretical frame}

\begin{figure*}
\includegraphics[width=2\columnwidth]{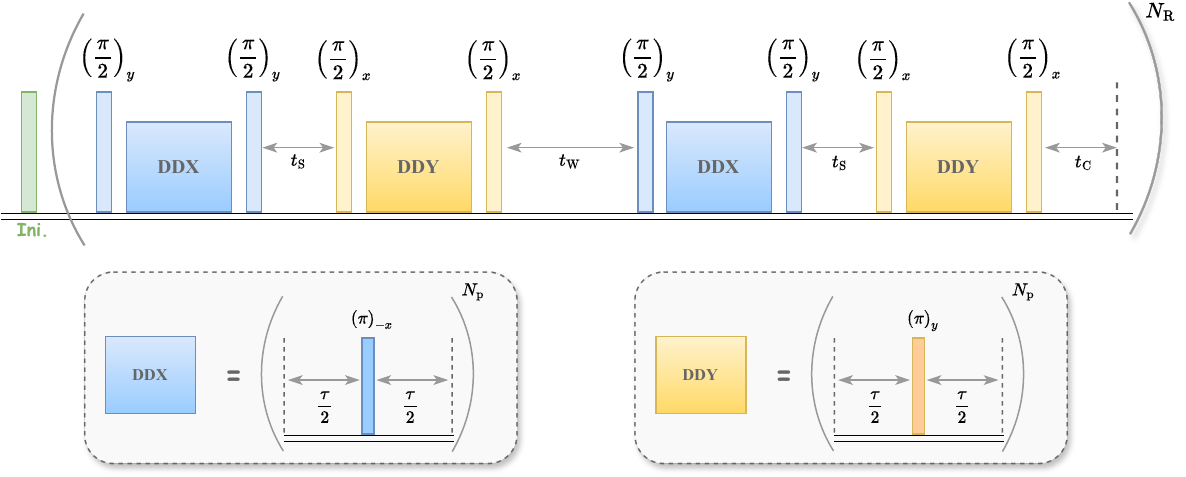} \caption{Hyper-polarization protocol of nuclear spins. The green box denotes
the initialization process of the electron spin. After the initialization,
the electron experiences two identical unitary processes($\mathrm{DDX}$+$\mathrm{DDY}$)
separated by a waiting time $t_{\mathrm{W}}$. There is also a delay
$t_{\mathrm{S}}$ between $\mathrm{DDX}$ and $\mathrm{DDY}$, which
represents the DD process under different $\pi$ pulses as shown in
the graphs below. Here $N_{\mathrm{p}}$ denotes the pulse number
of DD process. }
\label{Fig1} 
\end{figure*}

\subsection{Kraus operator form for hyper-polarization}

We consider that an electron spin(denoted by spin operator $\hat{\mathbf{S}}$)
interacts with the nuclear spin(denoted by spin operator $\hat{\mathbf{I}}$)
via the hyperfine coupling. The Hamiltonian can always be written
in the standard form 
\begin{equation}
\hat{H}=\omega\hat{I}_{z}+\hat{S}_{z}\mathbf{A}\cdot\hat{\mathbf{I}},\label{eq:Hamiltonian}
\end{equation}
in the rotation frame of electron spin when choosing a suitable coordinate
of nuclear spin. Here $\omega$ denotes the effective Larmor frequency
of the nuclear spin\cite{PfenderNC2019}. When choosing a proper $x$
axis of the nuclear spin, the hyperfine tensor can be written to $\mathbf{A}\equiv\{A_{\bot},0,A_{z}\}$
with $y$ component vanishing\cite{PfenderNC2019}.

The proposed Hyper-polarization protocol is shown in Fig.\ref{Fig1},
which includes $N$ units. In each unit, there are eight steps shown
as following: 
\begin{enumerate}
\item \textit{Electron spin initialization}: The electron spin is firstly
initialized to the state $|\Uparrow\rangle$, for example, by laser
illumination in NV center\cite{DohertyPR2013};\label{enu:initialization} 
\item DDX: DD sequence(with the $\pi$ pulse around $-x$ axis, see Fig.
\ref{Fig1}) with $\pi$ pulse number $N_{\mathrm{p}}$ and pulse
interval $\tau$, sandwiched by two half $\pi$ pulse around $y$
axis is applied. In principle, the axis of $\pi$ pulse can be any
direction in the $x-y$ plane;\label{enu:DDX} 
\item \textit{Waiting for a duration $t_{\mathrm{S}}$} : The nuclear spin
evolves freely without hyperfine coupling, which can be realized by
DD sequence with pulse interval unresonant with the nuclear spin\cite{ShenPRL2023};\label{enu:waiting-process} 
\item DDY: Another DD sequence(with the $\pi$ pulse around $y$ axis, see
Fig. \ref{Fig1}) sandwiched by two half $\pi$ pulse around $x$
axis is applied;\label{enu:DDY} 
\item \textit{Waiting for a duration $t_{\mathrm{W}}$ }: Another waiting
process with duration $t_{\mathrm{W}}$ is implemented. In this process,
the nuclear spin also evolves freely as the Step.\ref{enu:waiting-process}; 
\item Step. \ref{enu:DDX}-Step. \ref{enu:DDY} is repeated again; 
\item Another waiting time $t_{\mathrm{C}}$ to compensate for the phase
difference; \label{enu:Tcstep} 
\item Step.\ref{enu:DDX}-Step. \ref{enu:Tcstep} is repeated for $N_{\mathrm{R}}$
times. 
\end{enumerate}
The total unitary evolution of the total electron-nuclear spin system
is described by the unitary operation $\hat{U}$ (see details in the
Appendix. \ref{Appendix:Averaged_Hamiltonian}).

Then we use the unitary operation $\hat{U}$ to formulate the polarization
process clearly. After the initialization process(Step. \ref{enu:initialization}),
the density matrix of the nuclear spin and electron spin can always
be described by a separated state $\hat{\rho}=|\Uparrow\rangle\langle\Uparrow|\otimes\hat{\rho}_{\mathrm{N}}$
due to the initialization of electron spin. Denoting $\hat{\rho}_{\mathrm{N}}^{(n)}$
as the density matrix of the nuclear spin after the $n$th initialization,
the evolution of the nuclear spin state can be described by the recursion
equation 
\begin{equation}
\hat{\rho}_{\mathrm{N}}^{(n+1)}=\mathcal{M}\hat{\rho}_{\mathrm{N}}^{(n)},\label{eq:polar_dynamic}
\end{equation}
where the super-operator $\mathcal{M}$ is defined as 
\begin{equation}
\mathcal{M}\hat{\rho}_{\mathrm{N}}\equiv\mathrm{Tr}_{\mathrm{S}}\left[\hat{U}\left(|\Uparrow\rangle\langle\Uparrow|\otimes\hat{\rho}_{\mathrm{N}}\right)\hat{U}^{\dagger}\right],
\end{equation}
where $\mathrm{Tr}_{\mathrm{S}}$ denotes the trace over freedom of
electron spin. Inserting the completeness $|\Uparrow\rangle\langle\Uparrow|+|\Downarrow\rangle\langle\Downarrow|=1$
to the formula above, the super-operator $\mathcal{M}$ can be reformulated
to 
\begin{equation}
\mathcal{M}\hat{\rho}_{\mathrm{N}}=\hat{M}_{\Uparrow}\hat{\rho}_{\mathrm{N}}\hat{M}_{\Uparrow}^{\dagger}+\hat{M}_{\Downarrow}\hat{\rho}_{\mathrm{N}}\hat{M}_{\Downarrow}^{\dagger},\label{eq:Kraus}
\end{equation}
by using the Kraus operators $\hat{M}_{\Uparrow}=\langle\Uparrow|\hat{U}|\Uparrow\rangle$
and $\hat{M}_{\Downarrow}=\langle\Downarrow|\hat{U}|\Uparrow\rangle$
. Then the dynamics of the hyper-polarization of nuclear spin can
be simulated directly by Eq. (\ref{eq:polar_dynamic}).

\subsection{Approximated Analytical Result }

After the first order Magnus expansion, the evolution operator of
each unit $\hat{U}${[}Eq. (\ref{eq:Evolution_unit}){]} can be approximated
to the following standard form(see Appendix. \ref{Appendix:Averaged_Hamiltonian})

\begin{equation}
\hat{U}\approx-e^{-iN_{\mathrm{R}}\Phi\hat{I}_{z}}\exp\left\{ -i\alpha\left(\hat{S}_{x}\mathbf{e}_{\mathrm{X}}\cdot\hat{\mathbf{I}}+\hat{S}_{y}\mathbf{e}_{\mathrm{X},\phi}\cdot\hat{\mathbf{I}}\right)\right\} ,\label{eq:Average_hamilton}
\end{equation}
where the phase $\Phi=\omega T$ is precessing angle of the nuclear
spin during single unit of the $N_{\mathrm{R}}$ repetitions of the
protocol{[}Fig. \ref{Fig1}{]} and $T\equiv2t_{\mathrm{S}}+t_{\mathrm{W}}+4N_{\mathrm{p}}\tau+t_{\mathrm{C}}$
is corresponding duration. $\mathbf{e}_{\mathrm{X}},\mathbf{e}_{\mathrm{Y}}$
are two properly chosen unit vectors perpendicular to each other,
which depends on the detail of control parameters(see Appendix. \ref{Appendix:Averaged_Hamiltonian}).
$\mathbf{e}_{\mathrm{X},\phi}$ is a unit vector $\mathbf{e}_{\mathrm{X},\phi}=\cos\phi\mathbf{e}_{\mathrm{X}}+\sin\phi\mathbf{e}_{\mathrm{Y}}$
and $\phi$ is an angle depends on the time sequence 
\begin{equation}
\phi=\frac{(-1)^{N_{\mathrm{p}}}+1}{2}\pi-\omega\left(t_{\mathrm{S}}+N_{\mathrm{p}}\tau\right),\label{eq:phi}
\end{equation}
{[}modulus by $2\pi${]} and $\alpha$ is a real parameter defined
as following 
\begin{equation}
\alpha=2A_{\bot}\frac{\sin\left(N_{\mathrm{R}}\Phi/2\right)}{\sin\left(\Phi/2\right)}\sin\frac{\Phi_{1}}{2}F(\omega,N_{\mathrm{p}},\tau),\label{eq:alpha}
\end{equation}
whose absolute quantify the efficiency of the polarization protocol.
Here $\Phi_{1}=\omega\left(t_{\mathrm{S}}+t_{\mathrm{W}}+2N_{\mathrm{p}}\tau\right)$
is the precessing phase of the nuclear spin between the first DDX
and the second DDX and $F(\omega,N_{\mathrm{p}},\tau)$ is 
\[
F(\omega,N_{\mathrm{p}},\tau)=\begin{cases}
-\frac{4\sin\left(\frac{N_{\mathrm{p}}\omega\tau}{2}\right)\sin^{2}\left(\frac{\omega\tau}{4}\right)}{\omega\cos\left(\frac{\omega\tau}{2}\right)} & \mathrm{mod}\left(N_{\mathrm{p}},2\right)=0\\
\frac{4\cos\left(\frac{N_{\mathrm{p}}\omega\tau}{2}\right)\sin^{2}\left(\frac{\omega\tau}{4}\right)}{\omega\cos\left(\frac{\omega\tau}{2}\right)} & \mathrm{mod}\left(N_{\mathrm{p}},2\right)=1,
\end{cases}
\]
In Eq. (\ref{eq:alpha}), the first factor $\sin\left(N_{\mathrm{R}}\Phi/2\right)/\sin\left(\Phi/2\right)$
quantifies the synchronization between each evolution unit, the second
factor $\sin\left(\Phi_{1}/2\right)$ quantifies the synchronization
between the two DDX-DDY sequences in Fig. \ref{Fig1} and the third
factor $F(\omega,N_{\mathrm{p}},\tau)$ quantifies the resonance between
the DD pulse interval and the Larmor frequency of the nuclear spin\cite{YangRPP2017,PfenderNC2019}.

Using the approximation in Eq. (\ref{eq:Average_hamilton}), the Kraus
operator Eq. (\ref{eq:Kraus}) can be calculated analytically to be
\begin{equation}
\begin{alignedat}{1}\hat{M}_{\Uparrow}= & -\left(\begin{matrix}e^{-i\frac{\Phi}{2}}\cos\left(\alpha\cos\theta/2\right) & 0\\
0 & e^{i\frac{\Phi}{2}}\cos\left(\alpha\sin\theta/2\right)
\end{matrix}\right)\\
\hat{M}_{\Downarrow}= & \left(\begin{matrix}0 & -e^{-i\left(\theta+\frac{\Phi}{2}\right)}\sin\left(\alpha\sin\theta/2\right)\\
ie^{i\left(\theta+\frac{\Phi}{2}\right)}\sin\left(\alpha\cos\theta/2\right) & 0
\end{matrix}\right),
\end{alignedat}
\label{eq:Kraus_approx}
\end{equation}
in the basis of $|\uparrow\rangle$ and $|\downarrow\rangle$ of nuclear
spin basis and $\theta$ is an angle defined as $\theta=\phi/2+\pi/4$.
The Kraus operator in Eq. (\ref{eq:Kraus_approx}) can be used to
investigate analytically the polarization dynamics.

\section{Result}

\subsection{Optimal working point}

\begin{table*}
	\caption{The magic numbers for the Method.I and Method.II. The unit of the parameters $\tau$ and $t_{\mathrm{S}},t_{\mathrm{W}},t_{\mathrm{C}}$ are set to $\pi/\omega$. The left are the optimized parameters of Method.I while the right are that of Method.II. If $N_{\mathrm{R}}=1$, $t_{\mathrm{C}}=0$ while for other cases, $t_{\mathrm{C}}$ takes the value in this table. The polarization window $\Gamma${[}in unit of $\omega/(N_{\mathrm{R}}\pi)${]} and side bands $\delta\omega_{\mathrm{side}}$(in unit of $\pi/\omega$)of these two methods are also listed in this table.}
	\begin{tabular}{@{}cccccccccc@{}}
		\hline
		\hline
		&\multicolumn{3}{c}{Method.I}&\multicolumn{2}{c}{}&\multicolumn{3}{c}{Method.II}\\
		( $P_{\mathrm{s}}$, $N_{\mathrm{p}}$)  & $\tau${[}$\pi/\omega${]}  & $t_{\mathrm{S}},t_{\mathrm{W}},t_{\mathrm{C}}${[}$\pi/\omega$ mod($2\pi/\omega$){]}  & $\Gamma${[} $\frac{\omega}{N_{\mathrm{R}}\pi}${]}  & $\delta\omega_{\mathrm{side}}[\pi/\omega]$  & ( $P_{\mathrm{s}}$, $N_{\mathrm{p}}$)  & $\tau${[}$\pi/\omega${]}  & $t_{\mathrm{S}},t_{\mathrm{W}},t_{\mathrm{C}}${[}$\pi/\omega${]}  & $\Gamma${[} $\frac{\omega}{N_{\mathrm{R}}\pi}${]}  & $\delta\omega_{\mathrm{side}}${[}$\pi/\omega$ mod($2\pi/\omega$){]}\tabularnewline
		\hline 
		$+1,N_{\mathrm{p}}=1$  & $2$  & $3/2$  & $2/7$  & $\pm2k/7$  & $+1,N_{\mathrm{p}}=1$  & $3/2$  & $0$  & $1/3$  & $\pm2k/3$\tabularnewline
		$-1,N_{\mathrm{p}}=1$  & $2$  & $1/2$  & $2/5$  & $\pm2k/5$  & $-1,N_{\mathrm{p}}=1$  & $5/2$  & $0$  & $1/5$  & $\pm2k/5$\tabularnewline
		$+1,N_{\mathrm{p}}=2$  & $4/3$  & $11/6$  & $2/9$  & $\pm2k/9$  & $+1,N_{\mathrm{p}}=2$  & $5/4$  & $0$  & $2/5$  & $\pm2k/5$\tabularnewline
		$-1,N_{\mathrm{p}}=2$  & $4/3$  & $5/6$  & $2/7$  & $\pm2k/7$  & $-1,N_{\mathrm{p}}=2$  & $11/4$  & $0$  & $2/11$  & $\pm2k/11$\tabularnewline
		$+1$, $N_{\mathrm{p}}\ge3$  & $1$  & $1/2$  & $2/(2N_{\mathrm{p}}+1)$  & $\pm2k/(2N_{\mathrm{p}}+1)$  & $+1$, $N_{\mathrm{p}}\ge3$  & $1+\frac{1}{2N_{\mathrm{p}}}$  & $0$  & $2/(2N_{\mathrm{p}}+1)$  & $\pm2k/(2N_{\mathrm{p}}+1)$\tabularnewline
		$-1,$ $N_{\mathrm{p}}\ge3$  & $1$  & $3/2$  & $2/(2N_{\mathrm{p}}+3)$  & $\pm2k/(2N_{\mathrm{p}}+3)$  & $-1,$ $N_{\mathrm{p}}\ge3$  & $1-\frac{1}{2N_{\mathrm{p}}}$  & $0$  & $2/(2N_{\mathrm{p}}-1)$  & $\pm2k/(2N_{\mathrm{p}}-1)$\tabularnewline
		\hline
		\hline
	\end{tabular}
	\label{tab:Method_I_and_Method_II} 
\end{table*}

Before going to analyze in detail the performance from the Kraus operator
in Eq. (\ref{eq:Kraus_approx}), we first give an intuitive analysis
of stable polarization from the effective evolution operator in Eq.
(\ref{eq:Average_hamilton}). From Eq. (\ref{eq:Average_hamilton}),
the phase $\phi${[}Eq. (\ref{eq:phi}){]} is obviously an important
parameter which controls the stable polarization since it decides
the form of the effective Hamiltonian. To make the polarization degree
perfect, we need to set the phase $\phi$ to half integer of $\pi${[}see
Eq. (\ref{eq:phi}){]}, namely 
\begin{equation}
\phi=\begin{cases}
\frac{\pi}{2},\mathrm{perfect}\ \mathrm{positive}\ \mathrm{polarization}\\
\frac{3\pi}{2},\mathrm{perfect}\ \mathrm{negative}\ \mathrm{polarization}
\end{cases},\label{eq:phi_optimal}
\end{equation}
(modulus $2\pi$). For example, if $\phi$ is equal to $\pi/2$, then
we have $\mathbf{e}_{\mathrm{X},\phi}=\mathbf{e}_{\mathrm{Y}}$ and
hence exponent of Eq. (\ref{eq:Average_hamilton}) has form $\hat{S}_{+}\hat{I}_{-}+\hat{S}_{-}\hat{I}_{+}$,
which leads to full positive polarization. However, if the phase $\phi$
is tuned to $3\pi/2$, then we have $\mathbf{e}_{\mathrm{X},\phi}=-\mathbf{e}_{\mathrm{Y}}$
and hence exponent of Eq. (\ref{eq:Average_hamilton}) has form $\hat{S}_{+}\hat{I}_{+}+\hat{S}_{-}\hat{I}_{-}$,
which leads to full negative polarization.

To optimize the polarization performance, we need keep the perfect
polarization condition Eq. (\ref{eq:phi_optimal}) while simultaneously
maximizing $\alpha$ in Eq. (\ref{eq:alpha}). These parameters are
called optimal working point. There are two methods to achieve the
optimal working point. One method{[}called Method.I{]} is maximizing
$\alpha$ by tuning the three parameters $t_{\mathrm{S}},t_{\mathrm{W}},t_{\mathrm{C}}$
independently while simultaneously keeping the perfect polarization.
This is equal to maximizing the three factors of Eq. (\ref{eq:alpha})
respectively. So we need to set $\Phi$ to be $2k\pi$, $\Phi_{1}$
to be $(2k+1)\pi${[}see Eq. (\ref{eq:alpha}){]} and $\tau$ to be
resonant to nuclear spin{[}see Eq. (\ref{eq:alpha}) and Table. \ref{tab:resoanttau}
in the Appendix{]} while simultaneously fixing $\phi$ be half integer{[}Eq.
(\ref{eq:phi_optimal}){]}. The other method{[}called Method.II{]}
is maximizing $\alpha$ only by tuning $\tau$ when fixing $t_{\mathrm{S}}=t_{\mathrm{W}}=t_{\mathrm{C}}=0$
and simultaneously keeping the perfect polarization{[}$\phi$ be half
integer{[}Eq. (\ref{eq:phi_optimal}){]}. In this case, there is only
one parameter $\tau$ can be tuned. The magic parameters for Method.I
are shown in the left Table.\ref{tab:Method_I_and_Method_II} while
the Method.II in the right Table.\ref{tab:Method_I_and_Method_II}
after detailed analysis in the Appendix.\ref{subsec:First_method}
and Appendix.\ref{subsec:second_method}.

\subsection{Stable polarization}

\begin{figure}
\includegraphics[width=1\columnwidth]{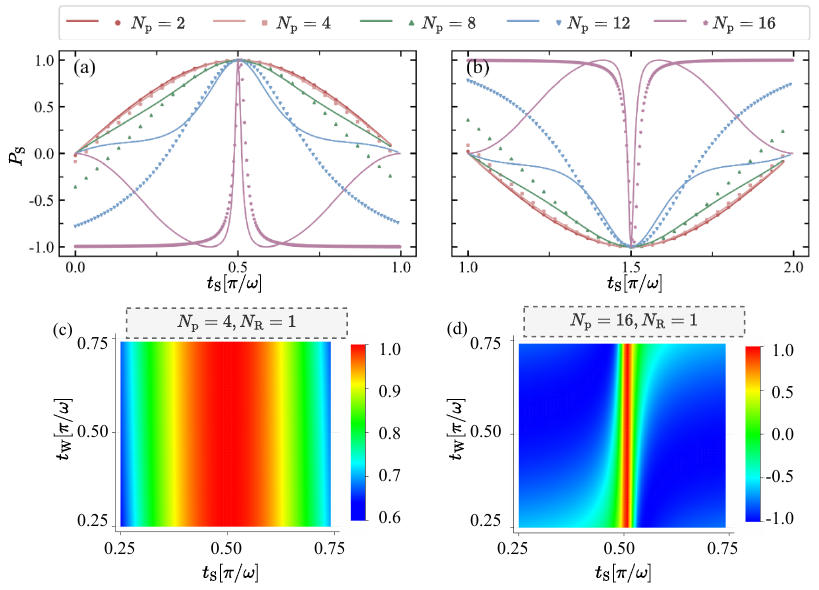} \caption{Stable polarization vs waiting time. The stable polarization as a
function of $t_{\mathrm{S}}$ for the case of (a) $N_{\mathrm{R}}=1$,
$t_{\mathrm{W}}=t_{\mathrm{C}}=\pi/(2\omega)$, $N_{\mathrm{p}}=2$(red),
$N_{\mathrm{p}}=4$(brown), $N_{\mathrm{p}}=8$(green), $N_{\mathrm{p}}=12$(blue)
and $N_{\mathrm{p}}=16$(purple). (b) $N_{\mathrm{R}}=1$, $t_{\mathrm{W}}=t_{\mathrm{C}}=3\pi/(2\omega)$,
$N_{\mathrm{p}}=2$(red), $N_{\mathrm{p}}=4$(brown), $N_{\mathrm{p}}=8$(green),
$N_{\mathrm{p}}=12$(blue) and $N_{\mathrm{p}}=16$(purple). The scatters
are the numerically simulated results while the lines with the same
color are the corresponding analytical results; The stable polarization
as a function of both $t_{\mathrm{S}},t_{\mathrm{W}}$ for the case
of (c) $N_{\mathrm{p}}=4,N_{\mathrm{R}}=1$ (d) $N_{\mathrm{p}}=16,N_{\mathrm{R}}=1$.
For all these graphs, the parameters of the Hamiltonian are $A_{\perp}=0.1,\omega=1$. }
\label{fig:stable_polar} 
\end{figure}

In the following, we analyze quantitatively the stable polarization
from the dynamical equation{[}Eq. (\ref{eq:polar_dynamic}){]} and
approximated Kraus operator{[}Eq. (\ref{eq:Kraus_approx}){]}. If
we only focus on the polarization, we can neglect the coherence of
the nuclear spin and the density matrix can be written as $\hat{\rho}_{\mathrm{N}}^{(N-1)}=P_{\uparrow}^{(N-1)}|\uparrow\rangle\langle\uparrow|+P_{\downarrow}^{(N-1)}|\downarrow\rangle\langle\downarrow|$
after the $N$th initialization($P_{\uparrow/\downarrow}^{(N)}$ is
the population of the nuclear state $|\uparrow/\downarrow\rangle$).
Inserting this equation to the Eq. (\ref{eq:polar_dynamic}) and using
the approximated expression of the Kraus operator Eq. (\ref{eq:Kraus_approx}),
we obtain the recursion equation for the nuclear spin population 
\begin{equation}
\left(\begin{array}{c}
P_{\uparrow}^{(N+1)}\\
P_{\downarrow}^{(N+1)}
\end{array}\right)=\left(\begin{matrix}\cos^{2}\left(\frac{\alpha\cos\theta}{2}\right) & \sin^{2}\left(\frac{\alpha\sin\theta}{2}\right)\\
\sin^{2}\left(\frac{\alpha\cos\theta}{2}\right) & \cos^{2}\left(\frac{\alpha\sin\theta}{2}\right)
\end{matrix}\right)\left(\begin{array}{c}
P_{\uparrow}^{(N)}\\
P_{\downarrow}^{(N)}
\end{array}\right).\label{eq:polar_equation}
\end{equation}
Using this equation, the stable polarization $P_{\mathrm{s}}\equiv P_{\uparrow}^{(+\infty)}-P_{\downarrow}^{(+\infty)}$
is calculated analytically to be

\begin{equation}
P_{\mathrm{s}}=\frac{\sin^{2}\left(\frac{1}{2}\alpha\sin\theta\right)-\sin^{2}\left(\frac{1}{2}\alpha\cos\theta\right)}{\sin^{2}\left(\frac{1}{2}\alpha\sin\theta\right)+\sin^{2}\left(\frac{1}{2}\alpha\cos\theta\right)},\label{eq:ana_Ps}
\end{equation}

The stable polarization can be controlled by the parameter $\phi$
from the analytical formula Eq. (\ref{eq:ana_Ps}) since $\theta=\phi/2+\pi/4$.
As shown in Eq. (\ref{eq:phi}), the parameter $\phi$ can be controlled
by the time delay $t_{\mathrm{S}}$ and DD parameters $N_{\mathrm{p}},\tau${[}Eq.
(\ref{eq:phi}){]} while is irrelevant to $t_{\mathrm{W}},t_{\mathrm{C}}$.
At the optimal working, $\phi$ is equal to half integer of $\pi$
and hence the stable polarization is perfect, which is independent
of the quantity $\alpha$. Specifically, for $\phi=\pi/2$($\theta=\pi/2$)(modulus
$2\pi$), Eq. (\ref{eq:ana_Ps}) gives perfect positive polarization.
For $\phi=3\pi/2$($\theta=\pi$)(modulus $2\pi$), however, Eq. (\ref{eq:ana_Ps})
gives perfect negative polarization, which is consistent with the
analysis from the form of the effective Hamiltonian in Eq. (\ref{eq:Average_hamilton}).

The stable polarization is highly sensitive to deviations in $t_{\mathrm{S}}$
when $\alpha$ is large. To validate these conclusions, we use the
parameters $\tau=\pi/\omega$, $t_{\mathrm{S}}=t_{\mathrm{W}}=t_{\mathrm{C}}=\pi/(2\omega)$
and $\tau=\pi/\omega$, $t_{\mathrm{S}}=t_{\mathrm{W}}=t_{\mathrm{C}}=3\pi/(2\omega)$
of Method.I in Table~\ref{tab:Method_I_and_Method_II}
to generate perfect positive and negative polarization, respectively.
By varying $t_{\mathrm{S}}$, we plot the stable polarization for
these two cases in Fig.~\ref{fig:stable_polar}(a) and (b) while
keep other parameters unchanged. Our results demonstrate that the
stable polarization varies significantly as a function of $t_{\mathrm{S}}$,
particularly for large $N_{\mathrm{p}}$. This behavior arises because
a large $\alpha$ enhances the sensitivity of stable polarization
to $\sin\theta$, and consequently to $t_{\mathrm{S}}$, as indicated
by the analytical formula in Eq.~\eqref{eq:ana_Ps}. Although Eq.~\eqref{eq:ana_Ps}
loses accuracy for large $\alpha$ when $\phi$ deviates from the
optimal working point defined in Eq.~\eqref{eq:phi_optimal}, it
still captures the qualitative trends of stable polarization as a
function of $t_{\mathrm{S}}$, as illustrated in Fig.~\ref{fig:stable_polar}(a)
and (b).

The stable polarization is insensitive to the deviation of the other
two waiting times $t_{\mathrm{W}},t_{\mathrm{C}}$. We plot the stable
polarization as a function of both $t_{\mathrm{S}},t_{\mathrm{W}}$
in Fig. \ref{fig:stable_polar} (c) $N_{\mathrm{p}}=4$ and (d) $N_{\mathrm{p}}=16$.
As shown in these figures, the stable polarization shows obvious dependence
on $t_{\mathrm{S}}$ while nearly independent on $t_{\mathrm{W}}$.
This can be roughly understood as following: When deviate from the
optimal working point, $|\alpha|$ becomes very small and hence we
can approximate the stable polarization as $P_{\mathrm{s}}\approx(\sin\theta-\cos\theta)/(\sin\theta+\cos\theta)$
from Eq. (\ref{eq:ana_Ps}). Consequently, the stable polarization
depends very weakly on $t_{\mathrm{W}},t_{\mathrm{C}}$ since $\theta$
is independent of $t_{\mathrm{W}},t_{\mathrm{C}}$.

\subsection{Polarization Rate}


The polarization rate of the nuclear spin is a key parameter to characterize
the performance of the hyper-polarization protocol. It can be obtained
by calculation of the dynamics of the polarization, which can be obtained
by solving the dynamics equation Eq. (\ref{eq:polar_equation}). The
result is 
\begin{equation}
P^{(N)}=P_{\mathrm{s}}(1-\lambda^{N-1}),\label{eq:polar_dynamic_ana}
\end{equation}
if the initial state of the nuclear spin is taken to the complete
mixed states(vanishing initial polarization). Here $\lambda$ 
\begin{equation}
\lambda=\frac{1}{2}\left|\cos\left(\alpha\sin\theta\right)+\cos\left(\alpha\cos\theta\right)\right|,\label{eq:lambda}
\end{equation}
is a quantity satisfying $\lambda\le1$. Under the optimal working
point($\theta=\pi/2$ and $\theta=\pi$), we have $\lambda=\cos^{2}(\alpha/2)$.
Using $\lambda$, we can define $N_{\mathrm{s}}=\max\left\{ -1/\ln\lambda,1\right\} $(the
number $1$ comes from the fact that the repetitive times can not
be smaller than $1$) to characterize the typical repetitive times
to polarize the nuclear spin(where $N_{\mathrm{s}}$ is defined as
$P^{(N_{\mathrm{s}})}/P_{\mathrm{s}}=1-e^{-1}$). Then the typical
polarization time can be defined as $T_{1}=N_{s}N_{\mathrm{R}}T$
and the polarization rate is then defined as $\gamma=1/T_{1}$ 
\begin{equation}
\gamma=\frac{\min\left\{ -\ln\lambda,1\right\} }{N_{\mathrm{R}}T}.\label{eq:polarization_rate}
\end{equation}

\begin{figure*}
\includegraphics[width=2\columnwidth]{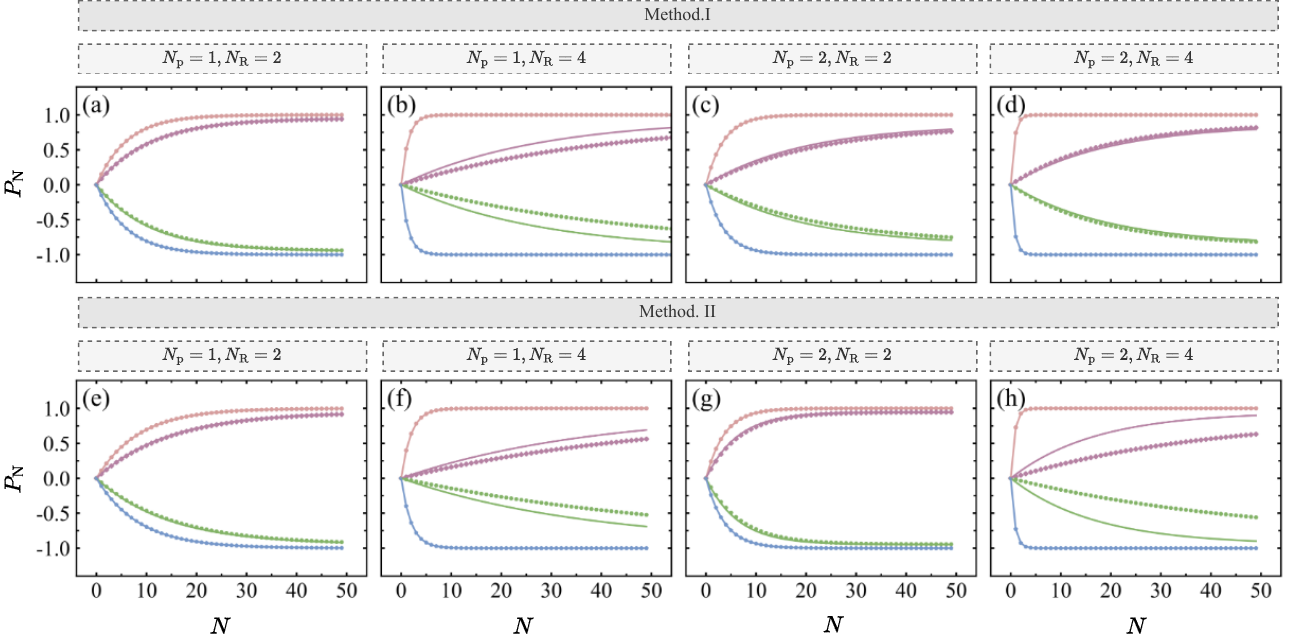} \caption{Polarization dynamics. $P_{N}$ as a function
of repetitive times $N$. The Scatters are the numerically simulated
results while the line with the same color are the corresponding analytical
result. The blue and pink scatters and line are the results for perfect
positive and negative polarization respectively with the parameters
taken from the left of Table. \ref{tab:Method_I_and_Method_II}. The
other cases are the parameters $t_{\mathrm{S}},t_{\mathrm{W}},t_{\mathrm{C}}$
with small deviation from these ideal parameters. Specifically, the
graphs above
plot the Method.I when (a) $N_{\mathrm{p}}=1,N_{\mathrm{R}}=2$. (b)
$N_{\mathrm{p}}=1,N_{\mathrm{R}}=4$. (c) $N_{\mathrm{p}}=2,N_{\mathrm{R}}=2$.
(d) $N_{\mathrm{p}}=2,N_{\mathrm{R}}=4$. $t_{\mathrm{S}}=t_{\mathrm{W}}=t_{\mathrm{C}}=2.8\pi/(2\omega)$(purple),
$t_{\mathrm{S}}=t_{\mathrm{W}}=t_{\mathrm{C}}=1.2\pi/(2\omega)$(green)
for $N_{\mathrm{p}}=1$ while $t_{\mathrm{S}}=t_{\mathrm{W}}=t_{\mathrm{C}}=10\pi/(6\omega)$(purple),
$t_{\mathrm{S}}=t_{\mathrm{W}}=t_{\mathrm{C}}=\pi/\omega$(green)
for $N_{\mathrm{p}}=2$. The  graphs
below plot the Method.II when (a) $N_{\mathrm{p}}=1,N_{\mathrm{R}}=2$.
(b) $N_{\mathrm{p}}=1,N_{\mathrm{R}}=4$. (c) $N_{\mathrm{p}}=2,N_{\mathrm{R}}=2$
and (d) $N_{\mathrm{p}}=2,N_{\mathrm{R}}=4$. The parameters are $t_{\mathrm{S}}=t_{\mathrm{W}}=t_{\mathrm{C}}=0$
for red and blue data while $t_{\mathrm{S}}=t_{\mathrm{W}}=t_{\mathrm{C}}=0.1\pi/\omega$
for the green and purple data. For all these figures, $\omega=1$
and $A_{\perp}=0.05,A_{z}=0$.}
\label{fig:Polarization_vs_time} 
\end{figure*}

To demonstrate the analytical result{[}Eq. (\ref{eq:polar_dynamic_ana}){]}
of the polarization dynamics, we compare it with that of the numerical
simulated result in Fig. \ref{fig:Polarization_vs_time} for the parameters
in Table. \ref{tab:Method_I_and_Method_II}. They are consistent with
each other very well for perfect polarization while only qualitatively
consistent when $\phi$ deviates from the optimal working point in
Table. \ref{tab:Method_I_and_Method_II}.

There is an approximated formula for polarization rate for weakly
coupled nuclear spin($A_{\perp}\ll\omega$) under optimal working
point. For weakly coupled nuclear spin, using the parameters in Table.
\ref{tab:Method_I_and_Method_II}, the polarization rate for the two
methods under the case of perfect polarization can be approximated
to 
\begin{equation}
\gamma_{\mathrm{opt}}\approx\frac{A_{\bot}}{\pi}\min\left\{ \frac{-\ln\cos\left(\alpha_{\mathrm{max}}/2\right)}{\alpha_{\mathrm{max}}/2},\frac{1}{\alpha_{\mathrm{max}}}\right\} ,\label{eq:gamma_approx}
\end{equation}
by noting that it requires large pulse number $N_{\mathrm{p}}$ for
weakly coupled case and hence the time cost in each unit is largely
spent in the duration of DD sequence. Here $\alpha_{\mathrm{max}}=4N_{\mathrm{R}}N_{\mathrm{p}}A_{\bot}/\omega$
is the maximal $\alpha$ which can be achieved for fixed $N_{\mathrm{R}}N_{\mathrm{p}}${[}see
Eq. (\ref{eq:alpha}){]}.

\begin{figure}
\includegraphics[width=1\columnwidth]{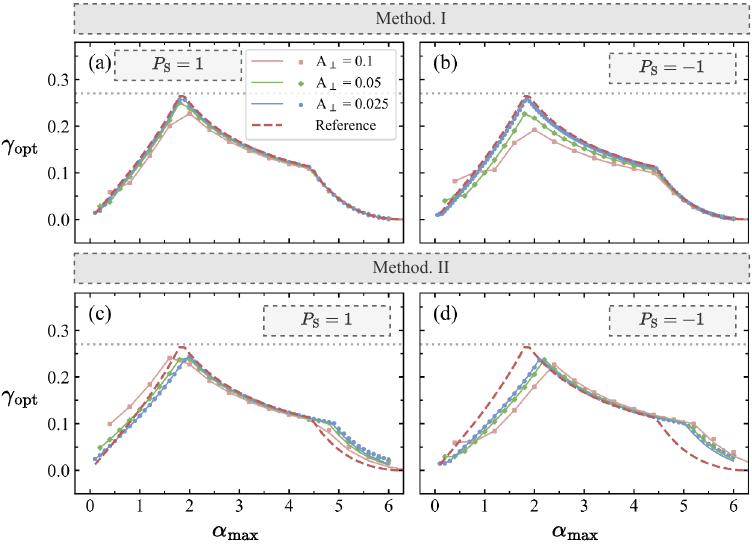} \caption{Polarization rate{[}in unit of $2A_{\bot}/\pi${]} vs $N_{\mathrm{p}}$.
$\gamma$ as a function of $\alpha_{\mathrm{max}}$ as $N_{\mathrm{p}}$
increases when\textbf{ }$N_{\mathrm{R}}=1$ is fixed. (a) Method.
I when $P_{\mathrm{s}}=1$. (b) Method.I
when $P_{\mathrm{s}}=-1$. (c) Method.II
when $P_{\mathrm{s}}=1$. (d) Method.II
method when $P_{\mathrm{s}}=-1$. The pink, green and blue data represent
the case of $A_{\bot}=0.1$, $A_{\bot}=0.05$ and $A_{\bot}=0.025$
respectively. The scatters are the numerically simulated results while
the line with the same color are the corresponding analytical results.
The red dashed linesis the reference line from the approximated formula
Eq. (\ref{eq:gamma_approx}) for weakly coupled limit. For all the
figures, the Larmor frequency is set to $\omega=1$.}
\label{fig:T1_vs_Np} 
\end{figure}

Under the optimal working point, the polarization rate shows an initial
linear increasing as the $N_{\mathrm{p}},N_{\mathrm{R}}$ increase
since the polarization rate $\gamma_{\mathrm{opt}}$ can be further
approximated to 
\begin{equation}
\gamma_{\mathrm{opt}}\approx\frac{A_{\bot}}{4\pi}\alpha_{\mathrm{max}}\approx N_{\mathrm{R}}N_{\mathrm{p}}\gamma_{0},\label{eq:gamma_opt_approx}
\end{equation}
for small $\alpha_{\mathrm{max}}$($\alpha_{\mathrm{max}}\ll1$).
Here $\gamma_{0}$ is defined as $A_{\bot}^{2}/(\pi\omega)$. 
To prove this relation, we simulate $\gamma_{\mathrm{opt}}$ as a
function of $\alpha_{\mathrm{max}}$ in Fig. \ref{fig:T1_vs_Np} for
both Method.I and Method.II under various parameters. In all these
figures, $\gamma_{\mathrm{opt}}$ shows an initial linear increasing
with respective to $\alpha_{\mathrm{max}}$. Besides, the consistent
between the analytical formula Eq.(\ref{eq:gamma_opt_approx}) and
the numerical result also enable us to measure the transverse coupling
$A_{\bot}$ via the polarization dynamics.

As the $N_{\mathrm{p}}$ continue increasing, the optimal polarization
rate will reach an universal maximal polarization rate $0.54A_{\bot}/\pi$
determined by the transverse hyperfine coupling $A_{\bot}$. From
Eq. (\ref{eq:gamma_approx}), the optimal polarization rate $\gamma_{\mathrm{opt}}$(in
unit of $2A_{\bot}/\pi$) is a sole function of $\alpha_{\mathrm{max}}$
and arrives its maximum\textbf{ $0.27$} when $\alpha_{\mathrm{max}}\approx1.84$.
As a result, the maximum of $\gamma_{\mathrm{opt}}$ is $0.54A_{\bot}/\pi$
limited by the hyperfine coupling $A_{\bot}$. As shown in Fig. \ref{fig:T1_vs_Np},
we plot $\gamma_{\mathrm{opt}}$ (in unit of $2A_{\bot}/\pi$) for
both two methods as a function of $\alpha_{\mathrm{max}}$ as the
$N_{\mathrm{p}}$ is changed. As the coupling $A_{\bot}$ decreases
from $A_{\bot}=0.1$(pink scatters in Fig. \ref{fig:T1_vs_Np}) to
$A_{\bot}=0.025$(blue scatters in Fig. \ref{fig:T1_vs_Np}), the
numerical result of $\gamma_{\mathrm{opt}}$ gradually converges to
reference line set by the approximated formula{[}Eq. (\ref{eq:gamma_approx}){]}
and shows an universal maximum $0.27$(in unit of $2A_{\bot}/\pi$)
for both the Method.I and Method.II.

\subsection{The effect of finite pulse duration}

In realistic polarization protocol, the duration of the half $\pi$
pulse can never be zero{[}see Fig. \ref{Fig1}{]} and is usually limited
by the micro wave power. As a result, we need consider how the duration
of the half $\pi$ pulse affects the polarization performance of the
methods in Table. \ref{tab:Method_I_and_Method_II}. For realistic
polarization protocol in NV center, the typical pulse duration $\tau_{\mathrm{\pi}}$
of the $\pi$ pulse is about $50\mathrm{ns}$(duration
of half $\pi$ pulse is $\tau_{\mathrm{\pi}}/2$) and the typical
Larmor frequency is about $\omega\sim2\pi\times1\mathrm{MHz}$ for
$^{13}\mathrm{C}$\cite{PfenderNC2019,ShenPRL2023} and $\omega\sim2\pi\times4\mathrm{MHz}$
for $^{1}\mathrm{H}$ under the magnetic field $0.1\mathrm{T}$. Consequently,
the maximum $\tau_{\mathrm{\pi}}$ can reach up to $0.4\pi/\omega$
and hence can not been neglected under strong
magnetic field or for nuclear spin with large gyromagnetic ratio.

\begin{figure}
\includegraphics[width=1\columnwidth]{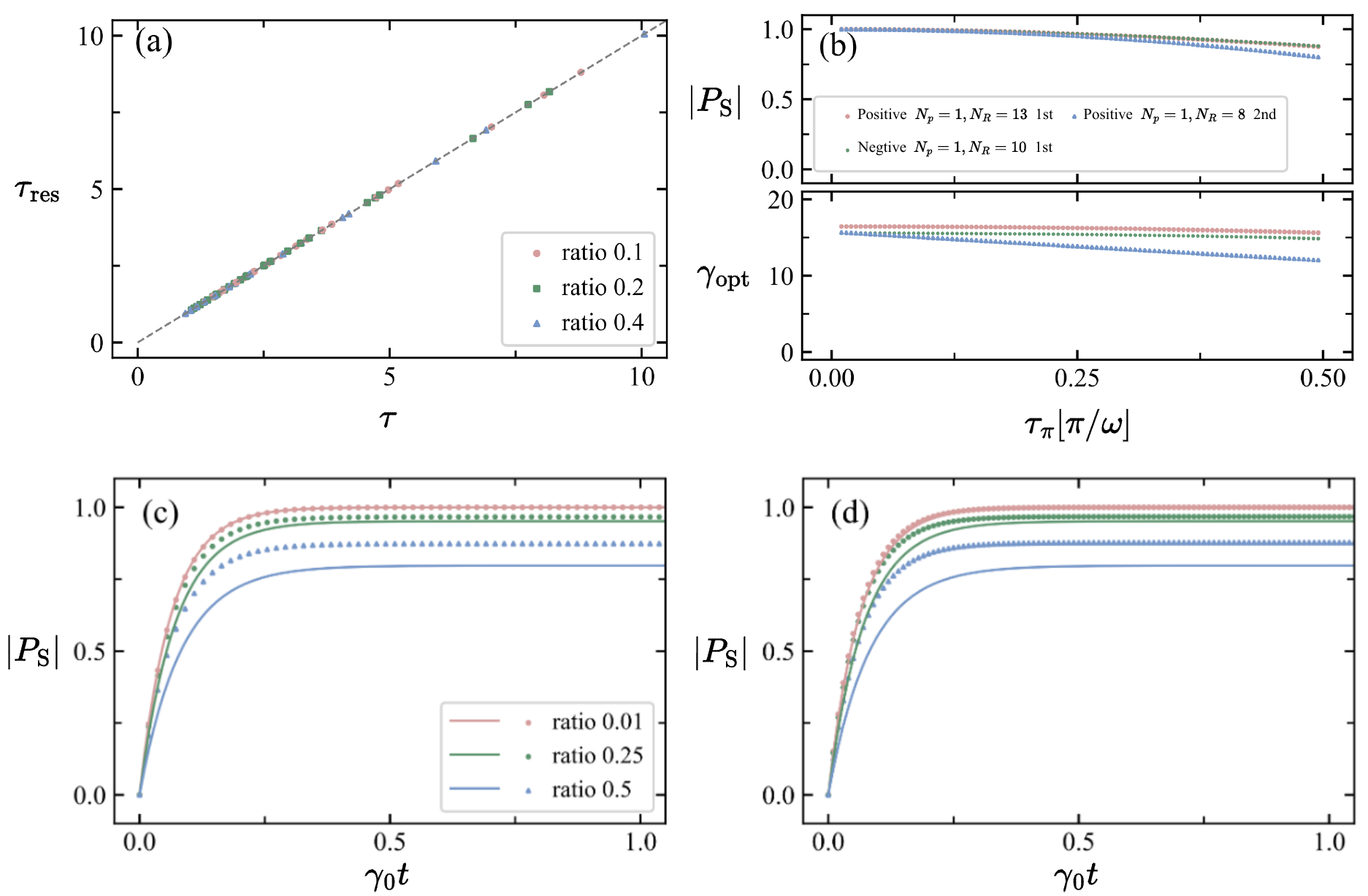} \caption{The effect of finite pulse duration. (a) The numerically searched
$\tau_{\mathrm{res}}$ maximizing the polarization rate as a function
of theoretical value $\tilde{\tau}$ modified by finite pulse duration
as shown in Eq. (\ref{eq:modefied_tau}). The black dashed line is
the reference line $\tau_{\mathrm{res}}=\tilde{\tau}$. Numerical
results are calculated for Method.I($N_{\mathrm{p}}=1,N_{\mathrm{R}}=8$
and $N_{\mathrm{p}}=2,N_{\mathrm{R}}=4$) and Method.II($N_{\mathrm{p}}=1,N_{\mathrm{R}}=8$)
under the case of $\tau_{\mathrm{\pi}}=0.1\pi/\omega${[}red dot{]},
$0.2\pi/\omega${[}green dot{]},$0.4\pi/\omega${[}blue dot{]} when
$\omega$ is changed from $0.5$ to $4$. (b) The absolute value of
the stable polarization(graph above) and polarization rate $\gamma$(graph
below) as a function of pulse duration $\tau_{\mathrm{\pi}}${[}in
unit of $\pi/\omega${]}. The parameters are: red dot(positive polarization
case of Method.I when $N_{\mathrm{p}}=1,N_{\mathrm{R}}=13$ in Table.
\ref{tab:Method_I_and_Method_II}), green dot(negative polarization
case of Method.I when $N_{\mathrm{p}}=1,N_{\mathrm{R}}=10$ in Table.
\ref{tab:Method_I_and_Method_II}) and blue dot(positive polarization
case of Method.II when $N_{\mathrm{p}}=1,N_{\mathrm{R}}=8$ in Table.
\ref{tab:Method_I_and_Method_II}). The Larmor is all set to $\omega=1$.
(c), (d) The comparison of the polarization dynamics between Method.I
and PulsePol method(positive case of $N_{\mathrm{p}}=1$ of Method.II
in Table. \ref{tab:Method_I_and_Method_II}) when $\tau_{\mathrm{\pi}}=0.5\pi/\omega$.
The other parameters is the same to (b). In graph (c) and (d) we compare
the polarization dynamics of the first parameter ($N_{\mathrm{p}}=1,N_{\mathrm{R}}=13$),
the second parameter($N_{\mathrm{p}}=1,N_{\mathrm{R}}=10$ ) of Method.I
and that of PulsePol method respectively. Red, blue, green solid lines
are the results of PulsePol method while the corresponding dashed
line are that of Method.I. For all these figures, the coupling is
set to $A_{\perp}=0.01$. }
\label{fig:finite_pulse_effect} 
\end{figure}

Firstly, we investigate how the pulse duration affects the optimal
working point shown in Table. \ref{tab:Method_I_and_Method_II}. We
keep all the waiting times $t_{\mathrm{W}},t_{\mathrm{S}},t_{\mathrm{C}}$
unchanged, the modified $\tilde{\tau}$ should be shift to 
\begin{equation}
\tilde{\tau}=\tau_{\mathrm{ideal}}-\frac{\tau_{\mathrm{\pi}}}{N_{\mathrm{p}}},\label{eq:modefied_tau}
\end{equation}
due to the existence of finite duration of half $\pi$ pulse, where
$\tau_{\mathrm{ideal}}$ is the ideal value in Table.\ref{tab:Method_I_and_Method_II}.
This formula is testified by numerical simulation shown in Fig.\ref{fig:finite_pulse_effect}(a).
We simulate the finite pulse duration by adding a Rabi term with Rabi
frequency be $\Omega=\pi/\tau_{\mathrm{\pi}}$ and then numerically
calculate the $\tau=\tau_{\mathrm{res}}$ which maximized the polarization
rate $\gamma$. We plot the optimized $\tau_{\mathrm{res}}$ as a
function of $\tilde{\tau}$ for various $\omega$ and $\tau_{\mathrm{\pi}}$
in Fig.\ref{fig:finite_pulse_effect}(a). They are equal to each other
as shown by the reference line in Fig.\ref{fig:finite_pulse_effect}(a)
and hence the optimized value in Eq. (\ref{eq:modefied_tau}) is proved
to be correct.

Then we show the advantage of our new protocol over the PulsePol protocol
in Ref. \cite{SchwartzSA2018}. To do this, we compare the robustness
of various protocol in Table.\ref{tab:Method_I_and_Method_II} to
the duration $\tau_{\mathrm{\pi}}$ under the modified optimal working
point{[}Eq. (\ref{eq:modefied_tau}){]}. We focus on the the first
parameters $N_{\mathrm{p}}=1,t_{\mathrm{S}}=t_{\mathrm{W}}=t_{\mathrm{S}}=0$
of the Method.II in
Table. \ref{tab:Method_I_and_Method_II}{[}essentially the PulsePol
protocol in Ref. \cite{SchwartzSA2018}{]} and the first two parameters
of Method.I in Table. \ref{tab:Method_I_and_Method_II}.
To do a relative fair comparison, we tune the optimized polarization
rate to be nearly the same for these three parameters under infinite
short $\tau_{\mathrm{\pi}}$(infinite large Rabi
frequency). This leads to the parameters $N_{\mathrm{p}}=1,N_{\mathrm{R}}=13,t_{\mathrm{S}}=t_{\mathrm{W}}=t_{\mathrm{S}}=3\pi/(2\omega)$,
$N_{\mathrm{p}}=1,N_{\mathrm{R}}=10,t_{\mathrm{S}}=t_{\mathrm{W}}=t_{\mathrm{S}}=\pi/(2\omega)$
and $N_{\mathrm{p}}=1,N_{\mathrm{R}}=8,t_{\mathrm{S}}=t_{\mathrm{W}}=t_{\mathrm{S}}=0$.
Then we calculate the stable polarization degree $|P_{\mathrm{s}}|$
and polarization rate $\gamma_{\mathrm{opt}}$ under the optimal working
point as a function of $\tau_{\mathrm{\pi}}${[}in unit of $\pi/\omega${]}.
These results are shown in Fig. \ref{fig:finite_pulse_effect} (b).
We can see that both the two new methods{[}$N_{\mathrm{p}}=1,N_{\mathrm{R}}=13,t_{\mathrm{S}}=t_{\mathrm{W}}=t_{\mathrm{S}}=3\pi/(2\omega)$
and $N_{\mathrm{p}}=1,N_{\mathrm{R}}=10,t_{\mathrm{S}}=t_{\mathrm{W}}=t_{\mathrm{S}}=\pi/(2\omega)${]}
proposed in this paper show obvious robustness over the
PulsePol protocol in Ref. \cite{SchwartzSA2018}{[}the parameter $N_{\mathrm{p}}=1,N_{\mathrm{R}}=8,t_{\mathrm{S}}=t_{\mathrm{W}}=t_{\mathrm{S}}=0${]}.
We also plot the polarization dynamics of $|P_{\mathrm{s}}|$ for various pulse duration
$\tau_{\mathrm{\pi}}${[}in unit of $\pi/\omega${]}. We can see
the dynamics of both the two new methods show obvious
robustness to $\tau_{\mathrm{\pi}}$ over that of PulsePol. 

\subsection{The polarization window and side bands}

\begin{figure}
\includegraphics[width=1\columnwidth]{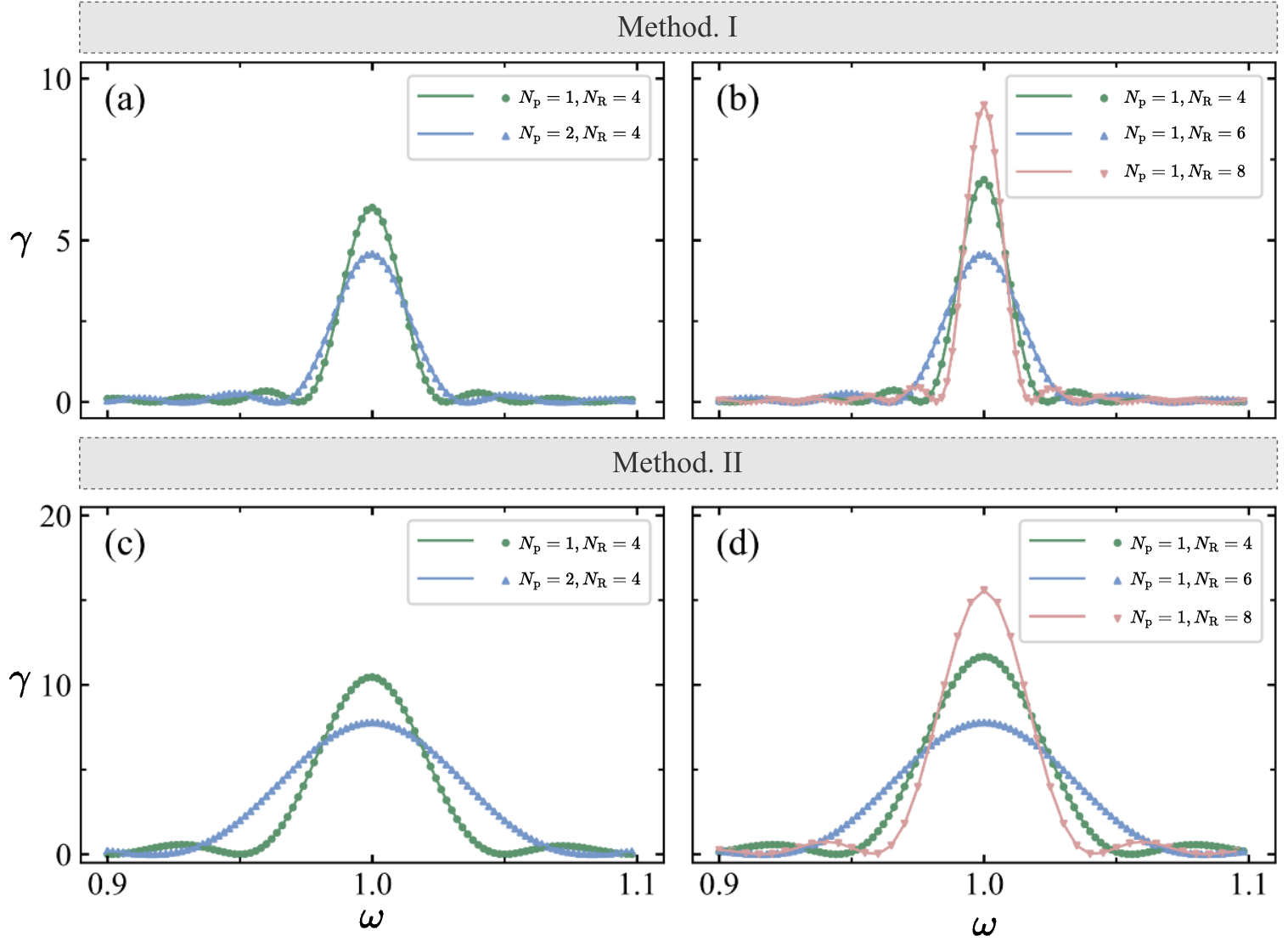} \caption{The polarization window for Method.I and Method.II. The polarization
rate{[}in unit of $\gamma_{0}\equiv A_{\bot}^{2}/(\pi\omega)${]}
as a function of $\omega$. The Method.I :\textcolor{magenta}{{} }(a)
Varying $N_{\mathrm{p}}$ while fixing $N_{\mathrm{R}}$; (b) Varying
$N_{\mathrm{R}}$ while fixing $N_{\mathrm{p}}$. The Method.II: (c)
Varying $N_{\mathrm{p}}$ while fixing $N_{\mathrm{R}}$. (d) Varying
$N_{\mathrm{R}}$ fixing $N_{\mathrm{p}}$ . For all the figures,
the Larmor frequency is set to $\omega=1$ and $A_{\perp}=0.005$.
The other parameters are taken from the Table.\ref{tab:Method_I_and_Method_II}
for different $N_{\mathrm{p}}$ and $N_{\mathrm{R}}$.}
\label{fig:rate_vs_omega} 
\end{figure}

In the following, we discuss the polarization window and side bands
of the polarization protocol. The polarization window $\delta\omega$
is defined as the frequency range in which the nuclear spins can be
effectively polarized, which quantified sensitivity of the protocol
to the error of Larmor frequency. While the side bands of the polarization
protocol is the extra polarization peak which will cause unwanted
polarization effect.

We find the polarization window is proportional $1/(N_{\mathrm{R}}T)$.
Since the stable polarization depends on $\phi$ in Eq. (\ref{eq:phi})
while the polarization rate depends on $\alpha$ in Eq. (\ref{eq:alpha}),
the stable polarization is less sensitive to the change of $\omega$
than the polarization rate if $N_{\mathrm{R}}$ is not equal to $1$.
As a result, the polarization window is determined by the sensitivity
of the polarization rate with respective to $\omega$, which is estimated
to be $\delta\omega=4/(N_{\mathrm{R}}T)$ at which the first factor
$\sin\left(N_{\mathrm{R}}\Phi/2\right)/\sin\left(\Phi/2\right)$ of
Eq. \ref{eq:alpha} decay to $\sin(2)/2\approx0.45$ of its maximum
value. As a result, the polarization window $\delta\omega$ can be
decreased by $N_{\mathrm{p}}$ and $N_{\mathrm{R}}$ as shown in Fig.
\ref{fig:rate_vs_omega} for both Method.I and Method.II. Once the
Larmor frequency $\delta\omega$ lies out this range, the polarization
rate will dramatically decrease. The polarization window $\delta\omega$
for different magic sequences are calculated in the Table. \ref{tab:Method_I_and_Method_II}.

\begin{figure}
\includegraphics[width=1\columnwidth]{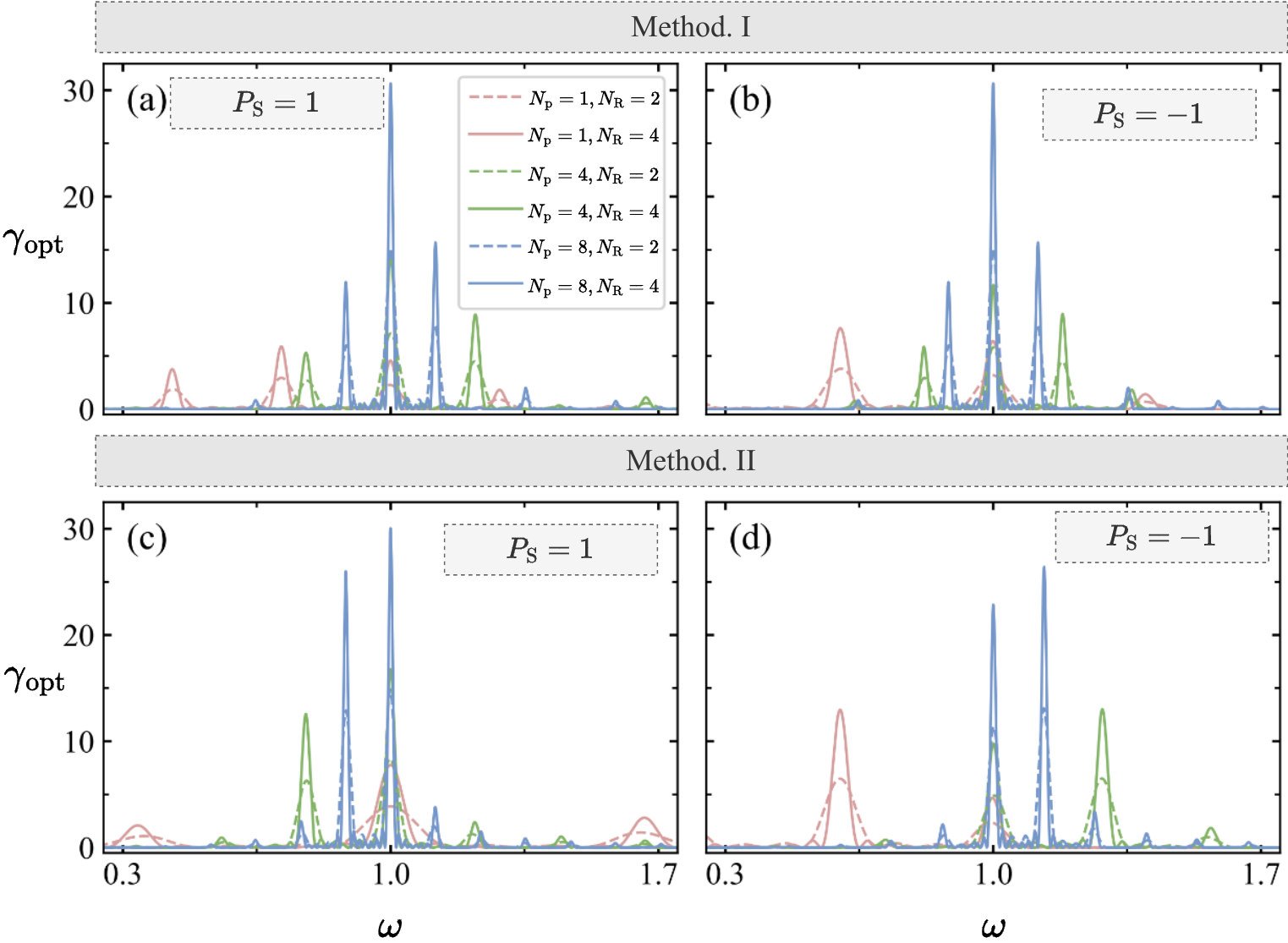} \caption{The side band of the polarization protocol. The polarization rate{[}in
unit of $\gamma_{0}=A_{\bot}^{2}/(\pi\omega)${]} is plot as a function
of Larmor frequency $\omega$ when other parameters is fixed. The
 panel above shows the result of the
Method.I when $P_{\mathrm{s}}$ is fixed to (a) $+1$ and (b) $-1$
for various pulse number $N_{\mathrm{p}}$ and $N_{\mathrm{R}}$,
which is shown in the legend of the figures. The 
panel below shows the result of the Method.II when $P_{\mathrm{s}}$
is fixed to (c) $+1$ and (d) $-1$ for the same pulse number $N_{\mathrm{p}}$
and $N_{\mathrm{R}}$. For all the figures, we set $\tau=\pi/\omega_{0}$
and $A_{\perp}=0.005,A_{z}=0$. The other parameters take the value
of Table. \ref{tab:Method_I_and_Method_II}.}
\label{fig:sideband} 
\end{figure}

The position of the first side bands highly depends on $N_{\mathrm{p}}$
while nearly independent of $N_{\mathrm{R}}$. If we denote the side
bands of the polarization rate by $\omega_{\mathrm{side}}$, the distant
$\delta\omega_{\mathrm{side}}=\omega_{\mathrm{side}}-\omega$ from
the central peak is calculated in the Table. \ref{tab:Method_I_and_Method_II}
from Eq. (\ref{eq:alpha}). In Fig. \ref{fig:sideband}, we plot the
polarization rate as a function of $\omega$ under various $N_{\mathrm{p}},N_{\mathrm{R}}$
for both the two methods. As $N_{\mathrm{p}}$ increases, the sides
bands will be closer to the central peak $\omega$ as shown in Fig.
\ref{fig:sideband}{[}see different colors and the same line shapes{]}.
However, the side bands is nearly unchanged when $N_{\mathrm{R}}$
is increased while keeping $N_{\mathrm{p}}$ invariant{[}see different
line shapes while the same colors in Fig. \ref{fig:sideband}{]}.
Since the side band will cause some unwanted effect, $N_{\mathrm{p}}$
should not be too large.

\section{Conclusion and Outlook}

In conclusion, we have developed a sequential protocol-based hyper-polarization
strategy for effectively polarizing weakly coupled nuclear spins.
Through systematic optimization, we have discovered a series of magic
control sequences that simultaneously maximize polarization degree
and polarization rate, establishing a digital framework for achieving
nuclear spin hyperpolarization in ensemble NV systems. Notably, our
protocol demonstrates superior robustness against the effects of finite
pulse duration compared to conventional methods like PulsePol, enabling
reliable operation across a broad range of magnetic field strengths
and nuclear spin species, particularly those with high gyro-magnetic
ratios(such as protons\cite{HealeyPRAppl2021}). This enhanced compatibility
with practical experimental conditions makes our approach particularly
suitable for integration with advanced NV center techniques, including
high-field single-shot readout methodologies\cite{NeumannScience2010}.
As a result, this compatibility of our protocol potentially enables
significant improvements in signal-to-noise ratios for NMR applications.
Besides, the general protocol is also applicable to other systems
such as radical system\cite{TanSA2019,WiliSA2022,RedrouthuJACS2022},
except for the NV centers. Further more, combination of two polarization
seqeuence resonant to two different nuclear spins can generate entanglement
between these nuclear spins and hence can be used to generate effective
strong coupling between nuclear spins, which may find important application
for quantum simulations\cite{RandallScience2021} in nuclear spins
system. 

\textit{Acknowledge---} P. W. is supported by National Natural Science
Foundation of China(Grant No. 12475012, Grant No. 62461160263), the
Guangdong Provincial Quantum Science Strategic Initiative (Grant GDZX2403009,
No. GDZX2303005), Innovation Program for Quantum Science and Technology
of China (Project 2023ZD0300600) and the Talents Introduction Foundation
of Beijing Normal University(Grant No. 310432106). H. L. is supported
by National Natural Science Foundation of China under Grant No. 62276171,
Guangdong Basic and Applied Basic Research Foundation, China, under
Grant No. 2024A1515011938, the Shenzhen Fundamental Research-General
Project, China under Grant No. JCYJ20240813141503005.

\appendix

\section{Deduction of equation(\ref{eq:Average_hamilton})}

\label{Appendix:Averaged_Hamiltonian}

As shown in Fig. \ref{Fig1}, the total system evolves in each unit
according to the following unitary operator 
\begin{eqnarray}
\hat{U} & = & \left(e^{-i\omega t_{\mathrm{C}}\hat{I}_{z}}\hat{U}_{1/2}e^{-i\omega t_{\mathrm{W}}\hat{I}_{z}}\hat{U}_{1/2}\right)^{N_{\mathrm{R}}},\label{eq:Evolution_unit}
\end{eqnarray}
after the initialization of electron spin. Here $\hat{U}_{1/2}$ is
defined as 
\begin{equation}
\hat{U}_{1/2}\equiv\hat{U}_{\mathrm{Y}}e^{-i\omega\hat{I}_{z}t_{\mathrm{S}}/2}\hat{U}_{\mathrm{X}},\label{eq:Uhalf}
\end{equation}
which denotes the unitary operator during the Step. \ref{enu:DDX}-Step.
\ref{enu:DDY}(DDX+waiting+DDY). Here 
\begin{equation}
\begin{alignedat}{1}\hat{U}_{\mathrm{Y}}= & \hat{U}_{\pi/2,x}\hat{U}_{\mathrm{DD},y}\hat{U}_{\pi/2,x}\\
\hat{U}_{\mathrm{X}}= & \hat{U}_{\pi/2,y}\hat{U}_{\mathrm{DD},-x}\hat{U}_{\pi/2,y},
\end{alignedat}
\label{eq:UXUY}
\end{equation}
denote the unitary evolution of Step. \ref{enu:DDX} and Step. \ref{enu:DDY},
where $\hat{U}_{\pi/2,x/y}$ is the half $\pi$ pulse around $x/y$
axis and $\hat{U}_{\mathrm{DD},y/(-x)}$ denotes the DD process with
$\pi$ pulse around $y/(-x)$ axis respectively. $\hat{U}_{\mathrm{DD},y/(-x)}$
can be formulated to 
\begin{align}
 & \hat{U}_{\mathrm{DD},-x}=\left(e^{-i\hat{H}\tau/2}\hat{U}_{\pi,-x}e^{-i\hat{H}\tau/2}\right)^{N_{\mathrm{p}}}\label{eq:UDD}\\
 & \hat{U}_{\mathrm{DD},y}=\left(e^{-i\hat{H}\tau/2}\hat{U}_{\pi,y}e^{-i\hat{H}\tau/2}\right)^{N_{\mathrm{p}}},
\end{align}
with $\hat{U}_{\pi,-x},\hat{U}_{\pi,y}$ denote the $\pi$ pulse around
$-x$ and $y$ axis respectively and $\hat{H}$ has been defined in
Eq. (\ref{eq:Hamiltonian}).

\subsection{Time ordered reformulation of $\hat{U}_{\mathrm{X}}$ and $\hat{U}_{\mathrm{Y}}$
in the main text}

In the following, we prove that the unitary evolution operators $\hat{U}_{\mathrm{X}}$
and $\hat{U}_{\mathrm{Y}}$ defined in Eq. (\ref{eq:Uhalf}) can be
reformulated as following

\begin{equation}
\begin{aligned}\hat{U}_{\mathrm{X}}= & -i\hat{\sigma}_{y}\mathrm{e}^{iN_{\mathrm{p}}\pi\hat{\sigma}_{z}/2}\times\mathrm{e}^{-i\omega\hat{I}_{z}N_{\mathrm{p}}\tau}\mathcal{T}\mathrm{e}^{i\int_{0}^{N_{\mathrm{p}}\tau}dtf_{\mathrm{DD}}\left(t\right)\hat{S}_{x}\mathbf{A}\cdot\hat{\mathbf{I}}\left(t\right)}\\
\hat{U}_{\mathrm{Y}}= & -i\hat{\sigma}_{x}\mathrm{e}^{iN_{\mathrm{p}}\pi\hat{\sigma}_{z}/2}\times\mathrm{e}^{-i\omega\hat{I}_{z}N_{\mathrm{p}}\tau}\mathcal{T}\mathrm{e}^{-i\int_{0}^{N_{\mathrm{p}}\tau}dtf_{\mathrm{DD}}\left(t\right)\hat{S}_{y}\mathbf{A}\cdot\hat{\mathbf{I}}\left(t\right)},
\end{aligned}
\label{eq:TO_UXUY}
\end{equation}
by moving all the $\pi$ and $\pi/2$ pulses to the beginning of the
evolution operators. Here $\mathcal{T}$ is the time ordered operator
and $\hat{\mathbf{I}}\left(t\right)\equiv e^{i\omega t\hat{I}_{z}}\hat{\mathbf{I}}e^{-i\omega t\hat{I}_{z}}$
is the time dependent spin operator. $f_{\mathrm{DD}}\left(t\right)=\pm1$
is the frequently used pulse shape function\cite{PfenderNC2019}.
An important feature of Eq. (\ref{eq:TO_UXUY}) is that $\hat{U}_{\mathrm{X}}$
only depends on $\hat{S}_{x}$ while $\hat{U}_{\mathrm{Y}}$ only
depends on $\hat{S}_{y}$.

Now, we prove Eq. (\ref{eq:TO_UXUY}). Using the identities $\hat{U}_{\pi,-x}^{\dagger}\hat{S}_{z}\hat{U}_{\pi,-x}=-\hat{S}_{z}$
and $\hat{U}_{\pi,y}^{\dagger}\hat{S}_{z}\hat{U}_{\pi,y}=-\hat{S}_{z}$,
we move $\hat{U}_{\pi,-x}$ and $\hat{U}_{\pi,y}$ of Eq. (\ref{eq:UXUY})
to the beginning of the operators and $\hat{U}_{\mathrm{DD},-x},\hat{U}_{\mathrm{DD},y}$
in Eq. (\ref{eq:UDD}) are reformulated to 
\begin{align}
 & \hat{U}_{\mathrm{DD},-x}=\left[\hat{U}_{\pi,-x}e^{-i\left(\omega\hat{I}_{z}-\hat{S}_{z}\mathbf{A}\cdot\hat{\mathbf{I}}\right)\tau/2}e^{-i\left(\omega\hat{I}_{z}+\hat{S}_{z}\mathbf{A}\cdot\hat{\mathbf{I}}\right)\tau/2}\right]^{N_{\mathrm{p}}}\nonumber \\
 & \hat{U}_{\mathrm{DD},y}=\left[\hat{U}_{\pi,y}e^{-i\left(\omega\hat{I}_{z}-\hat{S}_{z}\mathbf{A}\cdot\hat{\mathbf{I}}\right)\tau/2}e^{-i\left(\omega\hat{I}_{z}+\hat{S}_{z}\mathbf{A}\cdot\hat{\mathbf{I}}\right)\tau/2}\right]^{N_{\mathrm{p}}},
\end{align}
where the first $\hat{S}_{z}$ now obtains a negative sign. Then we
continue to move the $\pi$ pulse to the beginning of the all the
evolution operators and obtain 
\begin{align}
 & \hat{U}_{\mathrm{DD},-x}=\left(\hat{U}_{\pi,-x}\right)^{N_{\mathrm{p}}}\nonumber \\
 & \times\mathcal{T}\prod_{i=1}^{N_{\mathrm{p}}}\left[e^{-i\left[\omega\hat{I}_{z}+(-1)^{i}\hat{S}_{z}\mathbf{A}\cdot\hat{\mathbf{I}}\right]\tau/2}e^{-i\left[\omega\hat{I}_{z}-(-1)^{i}\hat{S}_{z}\mathbf{A}\cdot\hat{\mathbf{I}}\right]\tau/2}\right]\label{eq:UDDX1}
\end{align}
\begin{align}
 & \hat{U}_{\mathrm{DD},y}=\left(\hat{U}_{\pi,y}\right)^{N_{\mathrm{p}}}\nonumber \\
 & \times\mathcal{T}\prod_{i=1}^{N_{\mathrm{p}}}\left[e^{-i\left[\omega\hat{I}_{z}+(-1)^{i}\hat{S}_{z}\mathbf{A}\cdot\hat{\mathbf{I}}\right]\tau/2}e^{-i\left[\omega\hat{I}_{z}-(-1)^{i}\hat{S}_{z}\mathbf{A}\cdot\hat{\mathbf{I}}\right]\tau/2}\right],\label{eq:UDDY1}
\end{align}
where $\mathcal{T}$ is the time ordered operator with respective
to the index $i$. Introducing the pulse shape function 
\[
f_{\mathrm{DD}}\left(t\right)=\begin{cases}
1 & 0\le t<\tau/2\\
(-1)^{k} & (2k-1)\tau/2\le t<(2k+1)\tau/2\\
(-1)^{N_{\mathrm{p}}} & (N_{\mathrm{p}}-1/2)\tau\le t\le N_{\mathrm{p}}\tau
\end{cases},
\]
(where $k\in[1,N_{\mathrm{p}}-1]$), Eq. (\ref{eq:UDDX1}) and Eq.
(\ref{eq:UDDY1}) can be reformulated as the time ordered form 
\begin{align}
 & \hat{U}_{\mathrm{DD},-x}=\left(\hat{U}_{\pi,-x}\right)^{N_{\mathrm{p}}}e^{-i\omega\hat{I}_{z}N_{\mathrm{p}}\tau}\mathcal{T}e^{-i\int_{0}^{N_{\mathrm{p}}\tau}dtf_{\mathrm{DD}}\left(t\right)\hat{S}_{z}\mathbf{A}\cdot\hat{\mathbf{I}}\left(t\right)}\label{eq:TO_formu}\\
 & \hat{U}_{\mathrm{DD},y}=\left(\hat{U}_{\pi,y}\right)^{N_{\mathrm{p}}}e^{-i\omega\hat{I}_{z}N_{\mathrm{p}}\tau}\mathcal{T}e^{-i\int_{0}^{N_{\mathrm{p}}\tau}dtf_{\mathrm{DD}}\left(t\right)\hat{S}_{z}\mathbf{A}\cdot\hat{\mathbf{I}}\left(t\right)}.
\end{align}
Inserting it to Eq. (\ref{eq:UXUY}) and then moving all the pulses
to the beginning of the evolution operator again, $\hat{U}_{\mathrm{Y}}$
and $\hat{U}_{\mathrm{X}}$ are simplified to Eq. (\ref{eq:TO_UXUY})
when using the identity 
\begin{align*}
 & \hat{U}_{\pi/2,y}\left(\hat{U}_{\pi,-x}\right)^{N_{\mathrm{p}}}\hat{U}_{\pi/2,y}=-i\hat{\sigma}_{y}e^{iN_{\mathrm{p}}\pi\hat{\sigma}_{z}/2}\\
 & \hat{U}_{\pi/2,x}\left(\hat{U}_{\pi,y}\right)^{N_{\mathrm{p}}}\hat{U}_{\pi/2,x}=-i\hat{\sigma}_{x}e^{iN_{\mathrm{p}}\pi\hat{\sigma}_{z}/2},
\end{align*}

\subsection{First order Magnus approximation of $\hat{U}_{1/2}$ in Eq. (\ref{eq:Uhalf})}

Since $\hat{U}_{1/2}$ is constructed from $\hat{U}_{\mathrm{X}},\hat{U}_{\mathrm{Y}}$,
we firstly give the first order Magnus approximation of $\hat{U}_{\mathrm{X}},\hat{U}_{\mathrm{Y}}$.
The time ordered evolution operator in $\hat{U}_{\mathrm{X}},\hat{U}_{\mathrm{Y}}${[}Eq.
(\ref{eq:TO_UXUY}){]} can be approximated to 
\[
\mathcal{T}e^{\pm i\int_{0}^{N_{\mathrm{p}}\tau}dtf_{\mathrm{DD}}\left(t\right)\hat{S}_{x/y}\mathbf{A}\cdot\hat{\mathbf{I}}\left(t\right)}\approx e^{\pm i\hat{S}_{x/y}\int_{0}^{N_{\mathrm{p}}\tau}dtf_{\mathrm{DD}}\left(t\right)A_{\perp}\hat{I}_{x}\left(t\right)},
\]
using the first order Magnus expansion\cite{MaPRApplied2016,PfenderNC2019}{[}namely,
$\mathcal{T}e^{\pm i\int_{0}^{t}du\hat{O}\left(u\right)}\approx e^{\pm i\int_{0}^{t}du\hat{O}\left(u\right)}${]}.
Here $\hat{I}_{x}\left(t\right)=\hat{I}_{x}\cos\omega t-\hat{I}_{y}\sin\omega t$
is the time dependent spin operator. It should be noted that time
independent term $A_{z}\hat{I}_{z}$ term vanishes in the equation
above due to the integration $\int_{0}^{N_{\mathrm{p}}\tau}dtf_{\mathrm{DD}}\left(t\right)=0$.
Using the integration identity{[}see Appendix. \ref{subsec:pulse_identity}{]}
\begin{alignat}{1}
 & \int_{0}^{N_{\mathrm{p}}\tau}dtf_{\mathrm{DD}}\left(t\right)\cos\omega t=F(\omega,N_{\mathrm{p}},\tau)\nonumber \\
 & \times\begin{cases}
\cos\frac{N_{\mathrm{p}}\omega\tau}{2} & \mathrm{mod}\left(N_{\mathrm{p}},2\right)=0\\
\sin\frac{N_{\mathrm{p}}\omega\tau}{2} & \mathrm{mod}\left(N_{\mathrm{p}},2\right)=1
\end{cases}\label{eq:cos}
\end{alignat}
\begin{align}
 & \int_{0}^{N_{\mathrm{p}}\tau}dtf_{\mathrm{DD}}\left(t\right)\sin\omega t=F(\omega,N_{\mathrm{p}},\tau)\nonumber \\
 & \times\begin{cases}
\sin\frac{N_{\mathrm{p}}\omega\tau}{2} & \mathrm{mod}\left(N_{\mathrm{p}},2\right)=0\\
-\cos\frac{N_{\mathrm{p}}\omega\tau}{2} & \mathrm{mod}\left(N_{\mathrm{p}},2\right)=1
\end{cases},\label{eq:sin}
\end{align}
where 
\[
F(\omega,N_{\mathrm{p}},\tau)=\begin{cases}
-\frac{4\sin\left(\frac{N_{\mathrm{p}}\omega\tau}{2}\right)\sin^{2}\left(\frac{\omega\tau}{4}\right)}{\omega\cos\left(\frac{\omega\tau}{2}\right)}, & \mathrm{mod}\left(N_{\mathrm{p}},2\right)=0\\
\frac{4\cos\left(\frac{N_{\mathrm{p}}\omega\tau}{2}\right)\sin^{2}\left(\frac{\omega\tau}{4}\right)}{\omega\cos\left(\frac{\omega\tau}{2}\right)}, & \mathrm{mod}\left(N_{\mathrm{p}},2\right)=1,
\end{cases}
\]
quantifies the filter effect of the DD sequence.

Using Eq. (\ref{eq:cos}) and Eq. (\ref{eq:sin}), we obtain 
\begin{align*}
 & A_{\bot}\int_{0}^{N_{\mathrm{p}}\tau}dtf_{\mathrm{DD}}\left(t\right)\hat{I}_{x}\left(t\right)\\
 & =\alpha_{\mathrm{DD}}\begin{cases}
\mathbf{e}_{-\frac{N_{\mathrm{p}}\omega\tau}{2}}\cdot\hat{\mathbf{I}}, & \mathrm{mod}\left(N_{\mathrm{p}},2\right)=0\\
\mathbf{e}_{\frac{\pi}{2}-\frac{N_{\mathrm{p}}\omega\tau}{2}}\cdot\hat{\mathbf{I}}, & \mathrm{mod}\left(N_{\mathrm{p}},2\right)=1,
\end{cases}
\end{align*}
where $\mathbf{e}_{\varphi}=\cos\varphi\mathbf{e}_{x}+\sin\varphi\mathbf{e}_{y}$
and $\alpha_{\mathrm{DD}}$ is a dimensionless parameter 
\begin{equation}
\alpha_{\mathrm{DD}}=A_{\bot}F(\omega,N_{\mathrm{p}},\tau),
\end{equation}

As a result, the time ordered integration in $\hat{U}_{\mathrm{X}},\hat{U}_{\mathrm{Y}}$
of Eq. (\ref{eq:TO_UXUY})\textcolor{magenta}{{} }is simplified to  
\begin{equation}
\mathcal{T}e^{\pm i\int_{0}^{N_{\mathrm{p}}\tau}dtf_{\mathrm{DD}}\left(t\right)\hat{S}_{x/y}\mathbf{A}\cdot\hat{\mathbf{I}}\left(t\right)}\approx\mathrm{exp}\left\{ \pm i\alpha_{\mathrm{DD}}\hat{S}_{x/y}\mathbf{e}\cdot\hat{\mathbf{I}}\right\} ,\label{eq:unitary_Magnus}
\end{equation}
where $\mathbf{e}$ is an unit vector in the x-y plane 
\[
\mathbf{e}=\begin{cases}
\mathbf{e}_{-\frac{N_{\mathrm{p}}\omega\tau}{2}}, & \mathrm{mod}\left(N_{\mathrm{p}},2\right)=0\\
\mathbf{e}_{\frac{\pi}{2}-\frac{N_{\mathrm{p}}\omega\tau}{2}}, & \mathrm{mod}\left(N_{\mathrm{p}},2\right)=1,
\end{cases}
\]
Inserting Eq. (\ref{eq:unitary_Magnus}) to Eq. (\ref{eq:TO_UXUY}),
$\hat{U}_{\mathrm{X}}$ and $\hat{U}_{\mathrm{Y}}$ are approximated
to 
\begin{equation}
\begin{aligned}\hat{U}_{\mathrm{X}}\approx & -i\hat{\sigma}_{y}e^{iN_{\mathrm{p}}\pi\hat{\sigma}_{z}/2}\times e^{-i\omega\hat{I}_{z}N_{\mathrm{p}}\tau}\mathrm{e}^{+i\alpha_{\mathrm{DD}}\hat{S}_{x}\mathbf{e}\cdot\hat{\mathbf{I}}}\\
\hat{U}_{\mathrm{Y}}\approx & -i\hat{\sigma}_{x}e^{iN_{\mathrm{p}}\pi\hat{\sigma}_{z}/2}\times e^{-i\omega\hat{I}_{z}N_{\mathrm{p}}\tau}\mathrm{e}^{-i\alpha_{\mathrm{DD}}\hat{S}_{y}\mathbf{e}\cdot\hat{\mathbf{I}}}
\end{aligned}
,\label{eq:UXUY-1}
\end{equation}
under the first order Magnus expansion.

Inserting Eq. (\ref{eq:unitary_Magnus}) to Eq. (\ref{eq:Uhalf}),
$\hat{U}_{1/2}$ can be approximated to 
\[
\begin{gathered}\hat{U}_{1/2}\approx\left(-i\hat{\sigma}_{x}e^{iN_{\mathrm{p}}\pi\hat{\sigma}_{z}/2}e^{-i\omega\hat{I}_{z}N_{\mathrm{p}}\tau}e^{-i\alpha_{\mathrm{DD}}\hat{S}_{y}\mathbf{e}\cdot\hat{\mathbf{I}}}\right)\\
\times e^{-i\omega t_{\mathrm{S}}\hat{I}_{z}}\left(-i\hat{\sigma}_{y}e^{iN_{\mathrm{p}}\pi\hat{\sigma}_{z}/2}e^{-i\omega\hat{I}_{z}N_{\mathrm{p}}\tau}e^{i\alpha_{\mathrm{DD}}\hat{S}_{x}\mathbf{e}\cdot\hat{\mathbf{I}}}\right)
\end{gathered}
.
\]
Moving all the pulses to the beginning of $\hat{U}_{1/2}$ and using
the identity $\left(e^{iN_{\mathrm{p}}\pi\hat{\sigma}_{z}/2}\right)^{\dagger}\hat{S}_{y}e^{iN_{\mathrm{p}}\pi\hat{\sigma}_{z}/2}=\left(-1\right)^{N_{\mathrm{p}}}\hat{S}_{y}$
, we obtain 
\begin{equation}
\hat{U}_{1/2}\approx i\hat{\sigma}_{z}(-1)^{N_{\mathrm{p}}+1}\mathrm{e}^{-i\omega\hat{I}_{z}\left(t_{\mathrm{S}}+2N_{\mathrm{p}}\tau\right)}\mathrm{e}^{-i\alpha_{\mathrm{DD}}\left(-1\right)^{N_{\mathrm{p}}}\hat{S}_{y}\mathbf{e}_{-\Phi_{0}}\cdot\hat{\mathbf{I}}}\mathrm{e}^{i\alpha_{\mathrm{DD}}\hat{S}_{x}\mathbf{e}\cdot\hat{\mathbf{I}}},\label{eq:Uhalf_approx}
\end{equation}
with the $\Phi_{0}=\omega\left(t_{\mathrm{S}}+N_{\mathrm{p}}\tau\right)$
be the precessed phase of the nuclear between first DDX and first
DDY in Fig.\ref{Fig1} and $\mathbf{e}_{-\Phi_{0}}$ has been defined
before.

\subsection{Approximation of final evolution operator $\hat{U}$}

The evolution operator for each unit can be written as $\hat{U}\equiv\left(\hat{U}_{1}\right)^{N_{\mathrm{\mathrm{R}}}}$with
\begin{equation}
\hat{U}_{1}\equiv e^{-i\omega t_{\mathrm{C}}\hat{I}_{z}}\hat{U}_{1/2}e^{-i\omega t_{\mathrm{W}}\hat{I}_{z}}\hat{U}_{1/2},\label{eq:U1}
\end{equation}
denotes the evolution in single unit of $N_{\mathrm{\mathrm{R}}}$
repetitions. Consequently, to obtain the first order Magnus approximation
of $\hat{U}$, we firstly work out the approximated expression of
$\hat{U}_{1}$ via the approximation of $\hat{U}_{1/2}$ in Eq. (\ref{eq:Uhalf_approx})
and then use $\hat{U}_{1}$ to construct the approximation of $\hat{U}$.

\subsubsection{Approximation of $\hat{U}_{1}$ in Eq. (\ref{eq:U1})}

Using Eq. (\ref{eq:Uhalf_approx}) and moving the pulses to the beginning
as done previously, we find $\hat{U}_{1}$ in Eq.(\ref{eq:U1}) can
be approximated to 
\begin{equation}
\begin{alignedat}{1}\hat{U}_{1}\approx & -e^{-i\Phi\hat{I}_{z}}e^{i\alpha_{\mathrm{DD}}\left(-1\right)^{N_{\mathrm{p}}}\hat{S}_{y}\mathbf{e}_{-\Phi_{1}-\Phi_{0}}\cdot\hat{\mathbf{I}}}e^{-i\alpha_{\mathrm{DD}}\hat{S}_{x}\mathbf{e}_{-\Phi_{1}}\cdot\hat{\mathbf{I}}}\\
\times & e^{-i\alpha_{\mathrm{DD}}\left(-1\right)^{N_{\mathrm{p}}}\hat{S}_{y}\mathbf{e}_{-\Phi_{0}}\cdot\hat{\mathbf{I}}}e^{i\alpha_{\mathrm{DD}}\hat{S}_{x}\mathbf{e}\cdot\hat{\mathbf{I}}},
\end{alignedat}
\label{eq:Uapprox}
\end{equation}
where the phase $\Phi=\omega T\equiv\omega\left(2t_{\mathrm{S}}+t_{\mathrm{W}}+4N_{\mathrm{p}}\tau+t_{\mathrm{C}}\right)$
denotes the precessing angle of the nuclear spin in single unit of
$N_{\mathrm{R}}$ repetitions in Fig. \ref{Fig1} and $\Phi_{1}=\omega\left(t_{\mathrm{W}}+t_{\mathrm{S}}+2N_{\mathrm{p}}\tau\right)$
denotes the phase difference of the nuclear spin between the first
DDX and the second DDX in Fig. \ref{Fig1}.

Then we do first order Magnus expansion to $\hat{U}_{1}$ of Eq. (\ref{eq:Uapprox})
again and $\hat{U}_{1}$ is approximated to 
\begin{equation}
\hat{U}_{1}\approx-e^{-i\Phi\hat{I}_{z}}e^{-i\alpha_{\mathrm{DD}}\hat{L}},
\end{equation}
where 
\begin{equation}
\hat{L}=\left[\hat{S}_{x}\left(\mathbf{e}_{-\Phi_{1}}-\mathbf{e}\right)\cdot\hat{\mathbf{I}}+\left(-1\right)^{N_{\mathrm{p}}+1}\hat{S}_{y}\left(\mathbf{e}_{-\Phi_{1}-\Phi_{0}}-\mathbf{e}_{-\Phi_{0}}\right)\cdot\hat{\mathbf{I}}\right].\label{eq:Loperator}
\end{equation}
Defining new $x,y$ axis $\mathbf{e}_{x}^{\prime}=\mathbf{e}$ and
$\mathbf{e}_{y}^{\prime}=\mathbf{e}_{\pi/2}$($\mathbf{e}_{x}^{\prime}\perp\mathbf{e}_{y}^{\prime}$)
and using the the identity 
\[
\mathbf{e}_{\varphi_{2}}-\mathbf{e}_{\varphi_{1}}=2\sin\left(\frac{\varphi_{2}-\varphi_{1}}{2}\right)\left(-\sin\frac{\varphi_{2}+\varphi_{1}}{2}\mathbf{e}_{x}^{\prime}+\cos\frac{\varphi_{2}+\varphi_{1}}{2}\mathbf{e}_{y}^{\prime}\right),
\]
we obtain the following 
\begin{align}
 & \mathbf{e}_{-\Phi_{1}-\Phi_{0}}-\mathbf{e}_{-\Phi_{0}}=2\left(-1\right)^{N_{\mathrm{p}}+1}\sin\frac{\Phi_{1}}{2}\mathbf{e}_{x,\frac{3\pi}{2}-\frac{\Phi_{1}}{2}+\phi}^{\prime}\label{eq:difference_vector}\\
 & \mathbf{e}_{-\Phi_{1}}-\mathbf{e}=2\sin\frac{\Phi_{1}}{2}\mathbf{e}_{x,\frac{3\pi}{2}-\frac{\Phi_{1}}{2}}^{\prime},\nonumber 
\end{align}
with $\phi$ be the phase defined in the Eq. (\ref{eq:phi}) of the
main text.

Defining a new $x,y$ axis $\mathbf{e}_{\mathrm{X}}^{\prime},\mathbf{e}_{\mathrm{Y}}^{\prime}$
again 
\begin{equation}
\begin{aligned}\mathbf{e}_{\mathrm{X}}^{\prime}= & \mathbf{e}_{x,\frac{3\pi}{2}-\frac{\Phi_{1}}{2}}^{\prime}\\
\mathbf{e}_{\mathrm{Y}}^{\prime}= & \mathbf{e}_{\mathrm{X},\pi/2}^{\prime},
\end{aligned}
\label{eq:ex-ey}
\end{equation}
in Eq. (\ref{eq:difference_vector}) and inserting Eq. (\ref{eq:difference_vector})
to Eq. (\ref{eq:Loperator}), we obtain 
\begin{equation}
\hat{U}_{1}\approx-\mathrm{e}^{-i\Phi\hat{I}_{z}}\mathrm{e}^{-i2\sin\frac{\Phi_{1}}{2}\alpha_{\mathrm{DD}}\left[\hat{S}_{x}\mathbf{e}_{\mathrm{X}}^{\prime}\cdot\hat{\mathbf{I}}+\hat{S}_{y}\mathbf{e}_{\mathrm{X,\phi}}^{\prime}\cdot\hat{\mathbf{I}}\right]}.
\end{equation}

\subsubsection{Approximation of $\hat{U}$}

So the evolution operator $\hat{U}\equiv\left(\hat{U}_{1}\right)^{N_{\mathrm{\mathrm{R}}}}$
can be written as 
\begin{equation}
\hat{U}\approx\left[-\mathrm{e}^{-i\Phi\hat{I}_{z}}\mathrm{e}^{-i2\sin\frac{\Phi_{1}}{2}\alpha_{\mathrm{DD}}\left(\hat{S}_{x}\mathbf{e}_{\mathrm{X}}^{\prime}\cdot\hat{\mathbf{I}}+\hat{S}_{y}\mathbf{e}_{\mathrm{X,\phi}}^{\prime}\cdot\hat{\mathbf{I}}\right)}\right]^{N_{\mathrm{\mathrm{R}}}}.\label{eq:U_total}
\end{equation}
Then Eq. (\ref{eq:U_total}) can be reformulated as the time ordered
form again 
\begin{align*}
 & \hat{U}\approx\left(-1\right)^{N_{\mathrm{\mathrm{R}}}}e^{-iN_{\mathrm{\mathrm{R}}}\Phi\hat{I}_{z}}\\
 & \times\mathcal{T}\prod_{n=0}^{N_{\mathrm{\mathrm{R}}}-1}\exp\left\{ -2i\sin\frac{\Phi_{1}}{2}\alpha_{\mathrm{DD}}\left(\hat{S}_{x}\mathbf{e}_{\mathrm{X},-n\Phi}^{\prime}\cdot\hat{\mathbf{I}}+\hat{S}_{y}\mathbf{e}_{\mathrm{X},\phi-n\Phi}^{\prime}\cdot\hat{\mathbf{I}}\right)\right\} ,
\end{align*}
where $\mathcal{T}$ is the time ordered operator with respective
to the index $n$. Using the first order Magnus expansion again, we
obtain 
\[
\hat{U}\approx\left(-1\right)^{N_{\mathrm{\mathrm{R}}}}e^{-iN_{\mathrm{\mathrm{R}}}\Phi\hat{I}_{z}}\exp\left\{ -i\alpha\left(\hat{S}_{x}\mathbf{e}_{\mathrm{X}}\cdot\hat{\mathbf{I}}+\hat{S}_{y}\mathbf{e}_{\mathrm{X},\phi}\cdot\hat{\mathbf{I}}\right)\right\} ,
\]
when using the identity 
\[
\sum_{n=0}^{N_{\mathrm{\mathrm{R}}}-1}\mathbf{e}_{-n\Phi}=\frac{\sin\left(N_{\mathrm{\mathrm{R}}}\Phi/2\right)}{\sin\left(\Phi/2\right)}\mathbf{e}_{-\left(N_{\mathrm{\mathrm{R}}}-1\right)\Phi/2},
\]
for any unit vector $\mathbf{e}$. Here $\mathbf{e}_{\mathrm{X}}$
is a new unit vector 
\[
\mathbf{e}_{\mathrm{X}}=\mathbf{e}_{\mathrm{X},-\left(N_{\mathrm{\mathrm{R}}}-1\right)\Phi/2}^{\prime},
\]
and hence $\alpha$ becomes 
\begin{equation}
\alpha=2A_{\bot}\frac{\sin\left(N_{\mathrm{\mathrm{R}}}\Phi/2\right)}{\sin\left(\Phi/2\right)}\sin\frac{\Phi_{1}}{2}f(\omega,N_{\mathrm{p}},\tau),
\end{equation}
which is just the Eq. (\ref{eq:alpha}) in the main text.

\subsection{The deduction of integration of Eq. (\ref{eq:cos}) and Eq. (\ref{eq:sin})}

\label{subsec:pulse_identity}

In this subsection, we prove the integration of Eq. (\ref{eq:cos})
and Eq. (\ref{eq:sin}) . Firstly we calculate the integration, 
\[
I_{N_{\mathrm{p}}}=\int_{0}^{N_{\mathrm{p}}\tau}dtf_{\mathrm{DD}}\left(t\right)e^{i\omega t},
\]
because Eq. (\ref{eq:cos}) and Eq. (\ref{eq:sin}) are its real and
imaginary \textcolor{magenta}{parts} respectively.

If $\mathrm{mod}\left(N_{\mathrm{p}},2\right)=0$($N_{\mathrm{p}}$
is even), the integration can be written as 
\[
\begin{aligned}I_{N_{\mathrm{p}}}= & \sum_{n=0}^{N_{\mathrm{p}}/2-1}\int_{2n\tau}^{2\left(n+1\right)\tau}due^{i\omega t}f_{\mathrm{DD}}\left(t\right)\\
= & \sum_{n=0}^{N_{\mathrm{p}}/2-1}e^{i\omega2n\tau}\left[\int_{0}^{\tau/2}e^{i\omega t}dt-\int_{\tau/2}^{3\tau/2}e^{i\omega t}du+\int_{3\tau/2}^{2\tau}e^{i\omega t}dt\right].
\end{aligned}
\]
Using the identity 
\begin{align*}
 & \int_{0}^{\tau/2}e^{i\omega t}dt-\int_{\tau/2}^{3\tau/2}e^{i\omega t}dt+\int_{3\tau/2}^{2\tau}e^{i\omega t}dt\\
 & =-\frac{16\sin^{3}\left(\frac{\omega\tau}{4}\right)\cos\left(\frac{\omega\tau}{4}\right)}{\omega}e^{i\omega\tau},
\end{align*}
and 
\[
\sum_{n=0}^{N_{\mathrm{p}}/2-1}e^{i2n\omega\tau}=\frac{\sin\left(\frac{N_{\mathrm{p}}}{2}\omega\tau\right)}{\sin\left(\omega\tau\right)}e^{i\left(\frac{N_{\mathrm{p}}}{2}-1\right)\omega\tau},
\]
we obtain 
\begin{equation}
I_{N_{\mathrm{p}}}=-\frac{4\sin\left(\frac{N_{\mathrm{p}}}{2}\omega\tau\right)\sin^{2}\left(\frac{\omega\tau}{4}\right)}{\cos\left(\frac{\omega\tau}{2}\right)\omega}e^{i\frac{N_{\mathrm{p}}}{2}\omega\tau}.\label{eq:evenDD}
\end{equation}

If $\mathrm{mod}\left(N_{\mathrm{p}},2\right)=1$($N_{\mathrm{p}}$
is odd), then the integration can be divided to two parts, one is
the contribution from the even pulse number $N_{\mathrm{p}}-1$ while
the other part comes from the Hahn echo sequence beginning from the
time $\left(N_{\mathrm{p}}-1\right)\tau$ to $N_{\mathrm{p}}\tau$,
namely 
\[
\begin{aligned}I_{N_{\mathrm{p}}}= & \sum_{n=0}^{\frac{N_{\mathrm{p}}-1}{2}-1}\int_{2n\tau}^{2\left(n+1\right)\tau}dte^{i\omega t}f\left(t\right)\\
+ & \int_{\left(N_{\mathrm{p}}-1\right)\tau}^{\left(N_{\mathrm{p}}-1/2\right)\tau}dte^{i\omega t}-\int_{\left(N_{\mathrm{p}}-1/2\right)\tau}^{N_{\mathrm{p}}\tau}dte^{i\omega t},
\end{aligned}
\]
The first part can be calculated using the Eq. (\ref{eq:evenDD})
while the other part can be calculated directly. After simplification,
it becomes 
\begin{equation}
I_{N_{\mathrm{p}}}=\frac{4\cos\left(\frac{N_{\mathrm{p}}}{2}\omega\tau\right)\sin^{2}\left(\frac{\omega\tau}{4}\right)}{i\omega\cos\left(\frac{\omega\tau}{2}\right)}e^{i\frac{N_{\mathrm{p}}}{2}\omega\tau}.\label{eq:oddDD}
\end{equation}
Taking the real and imaginary part of $I_{N_{\mathrm{p}}}$ in Eq.
(\ref{eq:evenDD}) and Eq. (\ref{eq:oddDD}), we obtain the integration
in Eq. (\ref{eq:cos}) and Eq. (\ref{eq:sin}).

\section{Kraus operator}

\label{Appendix:Kraus-operator}

The exponent $\hat{S}_{x}\mathbf{e}_{\mathrm{X}}\cdot\hat{\mathbf{I}}+\hat{S}_{y}\mathbf{e}_{\mathrm{X},\phi}\cdot\hat{\mathbf{I}}$
of Eq. (\ref{eq:Average_hamilton}) can be divided to two terms

\begin{equation}
\hat{S}_{x}\mathbf{e}_{\mathrm{X}}\cdot\hat{\mathbf{I}}+\hat{S}_{y}\mathbf{e}_{\mathrm{X},\phi}\cdot\hat{\mathbf{I}}=\hat{G}_{+}+\hat{G}_{-},\label{eq:Average_hamilton-1}
\end{equation}
where 
\[
\begin{aligned}\hat{G}_{+}= & -i\frac{\sin\theta}{2}e^{i\theta}\hat{S}_{+}\hat{I}_{-}+h.c\\
\hat{G}_{-}= & \frac{\cos\theta}{2}e^{-i\theta}\hat{S}_{+}\hat{I}_{+}+h.c
\end{aligned}
,
\]
where $\theta\equiv\phi/2+\pi/4$ has been defined in the main text.
$\hat{G}_{+}$ only couples $|\Uparrow\downarrow\rangle$ , $|\Downarrow\uparrow\rangle$
while $\hat{G}_{-}$ only couples $|\Uparrow\uparrow\rangle$ and
$|\Downarrow\downarrow\rangle$. As a result, in the basis of $|\Uparrow\downarrow\rangle$
, $|\Downarrow\uparrow\rangle$, $|\Uparrow\uparrow\rangle$ and $|\Downarrow\downarrow\rangle$,
$\hat{S}_{x}\mathbf{e}_{\mathrm{X}}\cdot\hat{\mathbf{I}}+\hat{S}_{y}\mathbf{e}_{\mathrm{X},\phi}\cdot\hat{\mathbf{I}}$
can be formulated as the $4\times4$ matrix as following 
\begin{equation}
\hat{S}_{x}\mathbf{e}_{\mathrm{X}}\cdot\hat{\mathbf{I}}+\hat{S}_{y}\mathbf{e}_{\mathrm{X},\phi}\cdot\hat{\mathbf{I}}=\begin{pmatrix}0 & -i\frac{\sin\theta}{2}e^{i\theta} & 0 & 0\\
i\frac{\sin\theta}{2}e^{-i\theta} & 0 & 0 & 0\\
0 & 0 & 0 & \frac{\cos\theta}{2}e^{-i\theta}\\
0 & 0 & \frac{\cos\theta}{2}e^{i\theta} & 0
\end{pmatrix}.
\end{equation}
Consequently, the exponent $\exp\left\{ -i\alpha\left(\hat{G}_{+}+\hat{G}_{-}\right)\right\} $
can be calculated to the block form 
\begin{equation}
\hat{U}\approx\begin{pmatrix}\hat{U}_{\mathrm{u}} & 0\\
0 & \hat{U}_{\mathrm{d}}
\end{pmatrix},\label{eq:Uana}
\end{equation}
with 
\begin{equation}
\begin{alignedat}{1}\hat{U}_{\mathrm{u}}= & \begin{pmatrix}-e^{i\Phi/2}\cos\chi & e^{i(\theta+\Phi/2)}\sin\chi\\
-e^{-i(\theta+\Phi/2)}\sin\chi & -e^{-i\Phi/2}\cos\chi
\end{pmatrix}\\
\hat{U}_{\mathrm{d}}= & \begin{pmatrix}-e^{-i\Phi/2}\cos\eta & ie^{-i(\theta+\Phi/2)}\sin\eta\\
ie^{i(\theta+\Phi/2)}\sin\eta & -e^{i\Phi/2}\cos\eta
\end{pmatrix}
\end{alignedat}
,
\end{equation}
where $\eta=\alpha\cos\theta/2$ and $\chi=\alpha\sin\theta/2$. Using
the formula of Eq. (\ref{eq:Uana}), the Kraus operator can be calculated
to Eq. (\ref{eq:Kraus_approx}) in the main text from Eq.(\ref{eq:Kraus}).

\section{The parameters for optimized polarization rate}

\label{Appendix:Optimal_para}

\subsection{The Method.I }

\label{subsec:First_method}

For this method, the perfect polarization condition requires $\phi$
to be half integer, namely 
\begin{equation}
\frac{(-1)^{N_{\mathrm{p}}}+1}{2}\pi-\omega\left(t_{\mathrm{S}}+N_{\mathrm{p}}\tau\right)=\begin{cases}
\frac{4k_{1}+1}{2}\pi, & P_{\mathrm{s}}=1\\
\frac{4k_{1}+3}{2}\pi, & P_{\mathrm{s}}=-1
\end{cases},\label{eq:phi_opt}
\end{equation}
where $k_{1}$ is an integer. Maximizing $\alpha$ requires that the
three factors $\sin\left(N_{\mathrm{R}}\Phi/2\right)/\sin\left(\Phi/2\right)$,
$\sin\left(\Phi_{1}/2\right)$ and $f(\omega,N_{\mathrm{p}},\tau)$
in Eq. (\ref{eq:alpha}) are all maximized. This requires that $\Phi=\omega\left(t_{\mathrm{C}}+t_{\mathrm{W}}+2t_{\mathrm{S}}+4N_{\mathrm{p}}\tau\right)$
and $\Phi_{1}=\omega\left(t_{\mathrm{S}}+t_{\mathrm{W}}+2N_{\mathrm{p}}\tau\right)$
to be 
\begin{align}
 & \omega\left(t_{\mathrm{C}}+t_{\mathrm{W}}+2t_{\mathrm{S}}+4N_{\mathrm{p}}\tau\right)=2k_{3}\pi\nonumber \\
 & \omega\left(t_{\mathrm{S}}+t_{\mathrm{W}}+2N_{\mathrm{p}}\tau\right)=\left(2k_{2}+1\right)\pi,\label{eq:Phi_opt}
\end{align}
and $\tau$ to be the value in the Table.\ref{tab:resoanttau}. Here
$k_{2},k_{3}\ge0$ are also integers. In the following, we discuss
the optimized parameters for odd and even pulse number respectively.

\begin{table}
\caption{The pulse interval $\tau$ to be resonant to the nuclear spin for different pulses number\cite{YangRPP2017}.}
\begin{tabular}{cccc}
\hline
\hline 
Pulse interval $\tau$/Pulse number $N_{\mathrm{p}}$  & $N_{\mathrm{p}}=1$  & $N_{\mathrm{p}}=2$  & $N_{\mathrm{p}}\ge3$\tabularnewline
\hline 
$\tau${[}unit. $\pi/\omega${]}  & $2$  & $4/3,8/3$  & $\sim1$\tabularnewline
\hline 
\hline
\end{tabular}
\label{tab:resoanttau} 
\end{table}

\subsubsection{The case of even $N_{\mathrm{p}}$}

For even $N_{\mathrm{p}}=2$, we choose $\tau=4\pi/(3\omega)$ from
Table. \ref{tab:resoanttau}. The condition Eq. (\ref{eq:phi_opt})
and Eq. (\ref{eq:Phi_opt}) reduce to 
\begin{equation}
\begin{alignedat}{1}t_{\mathrm{S}}= & \frac{\pi}{\omega}\begin{cases}
\frac{4k_{1}+1}{2}-\frac{8}{3}, & P_{\mathrm{s}}=1\\
\frac{4k_{1}+3}{2}-\frac{8}{3}, & P_{\mathrm{s}}=-1
\end{cases}\\
t_{\mathrm{W}}= & \frac{\pi}{\omega}\left(2k_{2}-\frac{13}{3}\right)-t_{\mathrm{S}}\\
t_{\mathrm{C}}= & \left(2k_{3}-\frac{32}{3}\right)\frac{\pi}{\omega}-t_{\mathrm{W}}-2t_{\mathrm{S}},
\end{alignedat}
\end{equation}
where we have introduced new integers $k_{1},k_{2},k_{3}$ which are
different from the previous integers $k_{1},k_{2},k_{3}$. So the
minimal $t_{\mathrm{S}},t_{\mathrm{W}}$ takes the following number
\[
t_{\mathrm{S}}=t_{\mathrm{W}}=t_{\mathrm{C}}=\frac{\pi}{\omega}\begin{cases}
\frac{11}{6}, & P_{\mathrm{s}}=1\\
\frac{5}{6}, & P_{\mathrm{s}}=-1
\end{cases},
\]
due to $t_{\mathrm{S}},t_{\mathrm{W}},t_{\mathrm{C}}\ge0$.

For other even pulse number $N_{\mathrm{p}}>2$, we have $\tau=\pi/\omega$
from Table. \ref{tab:resoanttau}. Using this resonant $\tau$, the
condition of Eq. (\ref{eq:phi_opt}) and Eq. (\ref{eq:Phi_opt}) reduce
to 
\begin{equation}
\begin{alignedat}{1}t_{\mathrm{S}}= & \begin{cases}
\frac{4k_{1}+1}{2}\frac{\pi}{\omega}, & P_{\mathrm{s}}=1\\
\frac{4k_{1}+3}{2}\frac{\pi}{\omega}, & P_{\mathrm{s}}=-1
\end{cases}\\
t_{\mathrm{W}}= & \left(2k_{2}+1\right)\frac{\pi}{\omega}-t_{\mathrm{S}}\\
t_{\mathrm{C}}= & 2k_{3}\frac{\pi}{\omega}-t_{\mathrm{W}}-2t_{\mathrm{S}}.
\end{alignedat}
\end{equation}
{[}Here $k_{1}$ is a new integer{]}. As a result, the minimal $t_{\mathrm{S}},t_{\mathrm{W}}$
is 
\[
t_{\mathrm{S}}=t_{\mathrm{W}}=t_{\mathrm{C}}=\frac{\pi}{\omega}\begin{cases}
\frac{1}{2}, & P_{\mathrm{s}}=1\\
\frac{3}{2}, & P_{\mathrm{s}}=-1
\end{cases}.
\]

\subsubsection{The case of odd $N_{\mathrm{p}}$}

For odd $N_{\mathrm{p}}$, the condition Eq. (\ref{eq:phi_opt}) and
Eq. (\ref{eq:Phi_opt}) reduces 
\begin{equation}
\begin{alignedat}{1}t_{\mathrm{S}}= & \begin{cases}
\frac{4k_{1}+3}{2}\frac{\pi}{\omega}-N_{\mathrm{p}}\tau, & P_{\mathrm{s}}=1\\
\frac{4k_{1}+1}{2}\frac{\pi}{\omega}-N_{\mathrm{p}}\tau, & P_{\mathrm{s}}=-1
\end{cases}\\
t_{\mathrm{W}}= & \left(2k_{2}+1\right)\frac{\pi}{\omega}-t_{\mathrm{S}}\\
t_{\mathrm{C}}= & 2k_{3}\pi-t_{\mathrm{W}}-2t_{\mathrm{S}}-4N_{\mathrm{p}}\tau.
\end{alignedat}
\label{eq:phi_Phi_odd}
\end{equation}
For the special case $N_{\mathrm{p}}=1$, we use $\tau=2\pi/\omega$
in Table. \ref{tab:resoanttau} and Eq. (\ref{eq:phi_Phi_odd}) and
the the minimal $t_{\mathrm{S}},t_{\mathrm{W}}$ is 
\begin{equation}
t_{\mathrm{S}}=t_{\mathrm{W}}=t_{\mathrm{C}}=\frac{\pi}{\omega}\begin{cases}
\frac{3}{2}, & P_{\mathrm{s}}=1\\
\frac{1}{2}, & P_{\mathrm{s}}=-1
\end{cases}.\label{eq:optpara_Np}
\end{equation}
For the other odd pulse number $N_{\mathrm{p}}>1$, we use $\tau=\pi/\omega$
in Table. \ref{tab:resoanttau} and the the minimal $t_{\mathrm{S}},t_{\mathrm{W}}$
becomes 
\begin{equation}
t_{\mathrm{S}}=t_{\mathrm{W}}=t_{\mathrm{C}}=\frac{\pi}{\omega}\begin{cases}
\frac{1}{2}, & P_{\mathrm{s}}=1\\
\frac{3}{2}, & P_{\mathrm{s}}=-1
\end{cases}.\label{eq:phi_Phi_odd_Np}
\end{equation}
These results are summarized in the left of Table. \ref{tab:Method_I_and_Method_II}.

\subsection{The Method.II}

\label{subsec:second_method}

This method maximizes the polarization rate by maximizing $\alpha$
while fixing $t_{\mathrm{S}}=t_{\mathrm{W}}=t_{\mathrm{C}}=0$ and
hence there is only one parameter $\tau$. In the following, we discuss
the case of even $N_{\mathrm{p}}$ and odd $N_{\mathrm{p}}$ respectively.

\subsubsection{The case of even $N_{\mathrm{p}}$}

For even $N_{\mathrm{p}}$, the condition of perfect polarization{[}Eq.
(\ref{eq:phi_optimal}){]} and maximizing the first factor of $\alpha$
in Eq. (\ref{eq:alpha}) requires the following constraints 
\begin{equation}
\begin{alignedat}{1}N_{\mathrm{p}}\tau= & \frac{\pi}{\omega}\begin{cases}
\frac{4k_{1}+1}{2}, & P_{\mathrm{s}}=1\\
\frac{4k_{1}+3}{2}, & P_{\mathrm{s}}=-1
\end{cases}\\
N_{\mathrm{p}}\tau= & \frac{2k_{2}+1}{2}\frac{\pi}{\omega},
\end{alignedat}
\label{eq:phi_Pi_2nd_choosing}
\end{equation}
respectively. Since the first condition of Eq. (\ref{eq:phi_Pi_2nd_choosing})
exactly satisfies the second condition of Eq. (\ref{eq:phi_Pi_2nd_choosing}){]}.
As a result, the two constraints reduce to only one constraint 
\begin{equation}
\tau=\frac{\pi}{N_{\mathrm{p}}\omega}\begin{cases}
\frac{4k_{1}+1}{2}, & P_{\mathrm{s}}=1\\
\frac{4k_{1}+3}{2}, & P_{\mathrm{s}}=-1
\end{cases},\label{eq:tau_second_method_evenNp}
\end{equation}
for even $N_{\mathrm{p}}$. To maximize $\alpha$, there are two cases: 
\begin{enumerate}
\item For $N_{\mathrm{p}}=2$, to maximize the $\alpha$, we choose the
integer $k_{1}$ to make the pulse interval $\tau$ closest to the
resonant condition $\tau=4\pi/3\omega$, $\tau=8\pi/3\omega$. There
are two solutions 
\begin{equation}
k_{1}=\begin{cases}
\left[\frac{13}{12}\right]\equiv1, & P_{\mathrm{s}}=1\\
\\\left[\frac{7}{12}\right]\equiv1, & P_{\mathrm{s}}=-1
\end{cases},k_{1}=\begin{cases}
\left[\frac{29}{12}\right]\equiv3, & P_{\mathrm{s}}=1\\
\\\left[\frac{23}{12}\right]\equiv2, & P_{\mathrm{s}}=-1
\end{cases},
\end{equation}
where the function $\left[x\right]$ denotes the integer closest to
$x$. The closest case occurs when 
\begin{equation}
k_{1}^{\mathrm{opt}}=\begin{cases}
\left[\frac{13}{12}\right]\equiv1, & P_{\mathrm{s}}=1\\
\\\left[\frac{23}{12}\right]\equiv2, & P_{\mathrm{s}}=-1
\end{cases}.
\end{equation}
Using this optimized $k_{1}^{\mathrm{opt}}$, the pulse interval is
calculated to be 
\begin{equation}
\tau=\frac{\pi}{\omega}\begin{cases}
\frac{5}{4}, & P_{\mathrm{s}}=1\\
\frac{11}{4}, & P_{\mathrm{s}}=-1
\end{cases},\label{eq:tau_second_method_Np2}
\end{equation}
from Eq. (\ref{eq:tau_second_method_evenNp}). 
\item For even $N_{\mathrm{p}}>2$, to maximize $\alpha$, we choose the
integer $k_{1}^{\mathrm{opt}}$ to make the pulse interval $\tau$
closest to the approximated resonant condition $\tau=\pi/\omega$
and obtain the solutions 
\begin{equation}
k_{1}^{\mathrm{opt}}=\begin{cases}
\left[\frac{2N_{\mathrm{p}}-1}{4}\right], & P_{\mathrm{s}}=1\\
\\\left[\frac{2N_{\mathrm{p}}-3}{4}\right], & P_{\mathrm{s}}=-1
\end{cases}.\label{eq:condition_unopt-2}
\end{equation}
Using this optimized $k_{1}^{\mathrm{opt}}$, the pulse interval is
calculated to be 
\[
\tau=\frac{\pi}{N_{\mathrm{p}}\omega}\begin{cases}
\frac{4\left[\frac{N_{\mathrm{p}}}{2}-\frac{1}{4}\right]+1}{2}, & P_{\mathrm{s}}=1\\
\frac{4\left[\frac{N_{\mathrm{p}}}{2}-\frac{3}{4}\right]+3}{2}, & P_{\mathrm{s}}=-1
\end{cases}.
\]
Due to the definition of $\left[x\right]$ and the $\mathrm{mod}(N_{\mathrm{p}},2)=0$,
we have $\left[N_{\mathrm{p}}/2-1/4\right]=N_{\mathrm{p}}/2$ and
$\left[N_{\mathrm{p}}/2-3/4\right]=N_{\mathrm{p}}/2-1$ and hence
$\tau$ can be simplified to 
\[
\tau=\frac{\pi}{\omega}\begin{cases}
1+\frac{1}{2N_{\mathrm{p}}}, & P_{\mathrm{s}}=1\\
1-\frac{1}{2N_{\mathrm{p}}}, & P_{\mathrm{s}}=-1
\end{cases}.
\]
\end{enumerate}

\subsubsection{The case of odd $N_{\mathrm{p}}$}

For odd $N_{\mathrm{p}}$, the condition of perfect polarization{[}Eq.
(\ref{eq:phi_optimal}){]} and maximizing the first factor of $\alpha$
in Eq. (\ref{eq:alpha}) requires the following constraints 
\begin{equation}
\tau=\frac{\pi}{N_{\mathrm{p}}\omega}\begin{cases}
\frac{4k_{1}+3}{2}, & P_{\mathrm{s}}=1\\
\frac{4k_{1}+1}{2} & P_{\mathrm{s}}=-1
\end{cases}.\label{eq:condition_unopt_odd}
\end{equation}
To maximize $\alpha$, there are two cases for odd $N_{\mathrm{p}}$: 
\begin{enumerate}
\item For $N_{\mathrm{p}}=1$, the integer $k_{1}^{\mathrm{opt}}$ makes
the pulse interval $\tau$ closest to the resonant condition $\tau=2\pi/\omega${[}see
Table.\ref{tab:resoanttau}{]} is 
\begin{equation}
k_{1}^{\mathrm{op}}=\begin{cases}
0, & P_{\mathrm{s}}=1\\
\\1, & P_{\mathrm{s}}=-1
\end{cases}.\label{eq:para_2nd_Np}
\end{equation}
\item For $N_{\mathrm{p}}\ge3$, the integer $k_{1}^{\mathrm{op}}$ makes
the pulse interval $\tau$ closest to the resonant condition $\tau=\pi/\omega${[}see
Table.\ref{tab:resoanttau}{]} is 
\begin{equation}
k_{1}^{\mathrm{op}}=\begin{cases}
\left[\frac{2N_{\mathrm{p}}-3}{4}\right], & P_{\mathrm{s}}=1\\
\\\left[\frac{2N_{\mathrm{p}}-1}{4}\right], & P_{\mathrm{s}}=-1
\end{cases}.\label{eq:para_2nd_odd}
\end{equation}
Using these optimized $k_{1}^{\mathrm{op}}$, the pulse interval is
calculated to be 
\[
\tau=\frac{\pi}{N_{\mathrm{p}}\omega}\begin{cases}
\frac{4\left[\frac{N_{\mathrm{p}}}{2}-\frac{3}{4}\right]+1}{2} & P_{\mathrm{s}}=1\\
\frac{4\left[\frac{N_{\mathrm{p}}}{2}-\frac{1}{4}\right]+3}{2} & P_{\mathrm{s}}=-1
\end{cases}.
\]
Due to the definition of $\left[x\right]$ and the $\mathrm{mod}(N_{\mathrm{p}}+1,2)=0$,
we have $\left[\frac{N_{\mathrm{p}}}{2}-\frac{3}{4}\right]=\left[\frac{N_{\mathrm{p}}-1}{2}-\frac{1}{4}\right]=\left(N_{\mathrm{p}}-1\right)/2$
and $\left[\frac{N_{\mathrm{p}}}{2}-\frac{1}{4}\right]=\left[\frac{N_{\mathrm{p}}-1}{2}+\frac{1}{4}\right]=\left(N_{\mathrm{p}}-1\right)/2$
and hence $\tau$ can be simplified to 
\[
\tau=\frac{\pi}{\omega}\begin{cases}
1-\frac{1}{2N_{\mathrm{p}}}, & P_{\mathrm{s}}=1\\
1+\frac{1}{2N_{\mathrm{p}}}, & P_{\mathrm{s}}=-1
\end{cases}.
\]
These results are summarized in the right of the Table. \ref{tab:Method_I_and_Method_II}
of the main text. 
\end{enumerate}

%

\end{document}